\documentclass[aps,prd,twocolumn,showpacs,superscriptaddress,groupedaddress,nofootinbib]{revtex4}
\usepackage{xcolor}
\usepackage{graphicx,color,bm}

\usepackage{amsmath,amssymb}
\usepackage{epstopdf}
\usepackage{ulem}
\usepackage{array}
\usepackage{verbatim}
\usepackage[utf8]{inputenc}
\usepackage{rotating}
\newcolumntype{K}[1]{>{\centering\arraybackslash}m{#1}}

\usepackage[colorlinks,linkcolor=blue,citecolor=blue]{hyperref}

\newcommand{\orcid}[1]{\href{https://orcid.org/#1}{\,\includegraphics[width=8px]{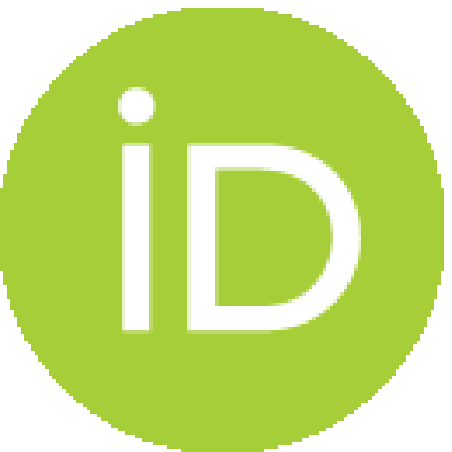}}}

\usepackage{mathtools}

\usepackage{tabularx}

\usepackage{tikz}
\usetikzlibrary{shapes.geometric, arrows}

\tikzstyle{startstop} = [rectangle, rounded corners, 
minimum width=3cm, 
minimum height=1cm,
text centered, 
draw=black, 
fill=red!30]

\tikzstyle{startstop2} = [rectangle, rounded corners, 
minimum width=1cm, 
minimum height=1cm,
text centered,
text width=1.3cm,
draw=black, 
fill=red!30]

\tikzstyle{startstop3} = [rectangle, rounded corners, 
minimum width=1cm, 
minimum height=1cm,
text centered,
text width=4cm,
draw=black, 
fill=red!30]

\tikzstyle{io} = [trapezium, 
trapezium stretches=true, 
trapezium left angle=70, 
trapezium right angle=110, 
minimum width=2cm, 
minimum height=1cm, text centered, 
text width=3.3cm,
draw=black, fill=blue!30]

\tikzstyle{process} = [rectangle, 
minimum width=3cm, 
minimum height=1cm, 
text centered, 
text width=3.6cm, 
draw=black, 
fill=orange!30]

\tikzstyle{process2} = [rectangle, 
minimum width=1cm, 
minimum height=1cm, 
text centered, 
text width=1.2cm, 
draw=black, 
fill=orange!30]

\tikzstyle{process3} = [rectangle, 
minimum width=1cm, 
minimum height=1cm, 
text centered, 
text width=1.8cm, 
draw=black, 
fill=orange!30]

\tikzstyle{process4} = [rectangle, 
minimum width=1cm, 
minimum height=1cm, 
text centered, 
text width=2.2cm, 
draw=black, 
fill=brown!30]

\tikzstyle{process5} = [rectangle, 
minimum width=1cm, 
minimum height=1cm, 
text centered, 
text width=1.9cm, 
draw=black, 
fill=orange!30]

\tikzstyle{decision} = [rectangle,
minimum width=1cm, 
minimum height=1cm, 
text centered,
text width=4cm,
draw=black, 
fill=green!30]
\tikzstyle{arrow} = [thick,->,>=stealth]

\tikzstyle{decision2} = [rectangle,
minimum width=1cm, 
minimum height=1cm, 
text centered,
text width=2.9cm,
draw=black, 
fill=yellow!30]
\tikzstyle{arrow} = [thick,->,>=stealth]

\tikzstyle{decision3} = [rectangle,
minimum width=1cm, 
minimum height=1cm, 
text centered,
text width=5cm,
draw=black, 
fill=black!30]
\tikzstyle{arrow} = [thick,->,>=stealth]

\begin{document}

\title{Model independent bounds on Type Ia supernova absolute peak magnitude}

\author{Bikash R. Dinda \orcid{0000-0001-5432-667X}}
\email{bikashdinda.pdf@iiserkol.ac.in}
\affiliation{ Department of Physical Sciences, Indian Institute of Science Education and Research Kolkata, India. }

\author{Narayan Banerjee \orcid{0000-0002-9799-2813}}
\email{narayan@iiserkol.ac.in}
\affiliation{ Department of Physical Sciences, Indian Institute of Science Education and Research Kolkata, India. }


\begin{abstract}
We put constraints on the peak absolute magnitude, $M_B$ of type Ia supernova using the Pantheon sample for type Ia supernova observations and the cosmic chronometers data for the Hubble parameter by a model independent and non-parametric approach. Our analysis is based on the Gaussian process regression. We find percent level bounds on the peak absolute magnitude given as $M_B=-19.384\pm0.052$. For completeness and to check the consistency of the results, we also include the Baryon acoustic oscillation data and the prior of the comoving sound horizon from Planck 2018 cosmic microwave background observations. The inclusion of these two data gives tighter constraints on $M_B$ at the sub-percent level. We obtain constraints on $M_B$ from the combination of pantheon compilation of type Ia supernova observations and baryon acoustic oscillation observations given as $M_B=-19.396\pm0.016$. When adding the cosmic chronometer observations with these observations, we find $M_B=-19.395\pm0.015$. The mean values of peak absolute magnitude from all these data are consistent with each other and the values are approximately equal to $-19.4$.
\end{abstract}


\maketitle
\date{\today}

\section{Introduction}
\label{sec-intro}

The late time cosmic acceleration was first discovered by the type Ia supernovae observations \citep{SupernovaSearchTeam:1998fmf,SupernovaCosmologyProject:1998vns,2011NatPh...7Q.833W}. These observations are based on the fact that the type Ia supernovae are standard candles and the peak absolute magnitude, $M_B$ of a type Ia supernova is uniform. The discovery of the late time cosmic acceleration led to the concept of dark energy (for details see \citep{Peebles:2002gy,SupernovaCosmologyProject:2008ojh}), where the dark energy is considered to be an exotic matter component in the Universe that has an effective large negative pressure.

The peak absolute magnitude, $M_B$ of type Ia supernova plays an important role in the determination of the expansion history of the Universe since the cosmic distances like the luminosity distance of an astronomical object are related to the distance modulus of the Type Ia supernovae. This distance modulus depends both on the observed magnitude, $m$, and the absolute magnitude, $M_B$ \citep{Linden2009CosmologicalPE,Camarena:2019rmj,Pan-STARRS1:2017jku,Camlibel:2020xbn}. That is why the observational constraints on the cosmological parameters like the deceleration parameter, the matter-energy density parameter, the dark energy density parameter, etc. are estimated based on the value of the absolute peak magnitude, $M_B$ \citep{Cao:2022ugh,Colgain:2022nlb}. Thus it is important to know the exact value of $M_B$.

In most of the recent type Ia supernova-based cosmological studies, the parameter $M_B$ is considered to be a nuisance parameter and fitted with the other parameters of the stretch color relation \citep{Tripp:1997wt}. This is because the constraints on the cosmological parameters from the type Ia supernova observations are degenerate to $M_B$ and this degeneracy stems from the degeneracy between $M_B$ and $H_0$ (present value of the Hubble parameter). This is the reason, alternatively, we need to calibrate $M_B$ by combining type Ia supernova data with other astrophysics and cosmological data \citep{Camarena:2019rmj}. We also note that, in recent investigations, the determination of the distance modulus from the type Ia supernova observations is dependent on the distance bias corrections, mass step corrections, etc \citep{Pan-STARRS1:2017jku}. 
Indeed, this steals the importance of $M_B$ a bit, but the value of $M_B$ still finds use in the determination of the Hubble constant, $H_0$ whenever type Ia supernova observations are considered \citep{Camarena:2019moy,Philcox:2022sgj}.

In literature, we find the inconsistency in the values of $H_0$ from low redshift observations like SHOES \citep{Riess:2020fzl} and the high redshift observations like cosmic microwave background (CMB) \citep{Planck:2018vyg}. This is the so-called Hubble tension \citep{DiValentino:2021izs,Vagnozzi:2019ezj,Krishnan:2021dyb}. However, recently, some authors have argued that the Hubble tension is not the fundamental tension, rather $M_B$ tension is the more fundamental one when we compare the low redshift observations with the high redshift observations (with the presence of the type Ia supernova observations). For details see \citep{Camarena:2021jlr,Efstathiou:2021ocp,Dinda:2021ffa}. Thus, in this regard, the determination of $M_B$ from different combinations of data is also important.

We also find the relevance of the knowledge of $M_B$ in some other cases. For example, for some non-standard cosmological studies like the measurement of Newtonian gravitational constant and its time variation \citep{Zhao:2018gwk}, the determination of the speed of light and its time variation \citep{Colaco:2022noc}, and the determination of the fine structure constant and also its time variation \citep{Colaco:2022noc,Colaco:2022aut} from the combination of supernova and other observations depend on the value of the type Ia supernova peak absolute magnitude. So, in these cases, the value of $M_B$ plays a key role. Note that these studies are independent of the degeneracy between $M_B$ and $H_0$. This is because these studies are directly dependent on the cosmological distance like luminosity distance. These are not directly dependent on the Hubble parameter or the relevant quantities. That is why $H_0$ is not involved. So, in these cases, the $M_B$ parameter is the main parameter and the results depend on its value.

Thus it is found that $M_B$ still plays a crucial role in some measurements and a secondary role in some others, and is still quite a relevant quantity. This motivates the present work, which deals with the estimation of $M_B$.

The determination of $M_B$ is based on the anchors like stellar parallax \citep{vanLeeuwen:2007xw,2018ApJ...855..136R,Greene:2021shv}, detached eclipsing binary stars \citep{Pietrzynski:2013gia}, and maser emission from supermassive black holes \citep{Reid:2019tiq,Pihlstrom:2004vg,Gao:2015tqd}. These methods are mainly astrophysical and restricted to lower redshift observations only. For example, in SHOES observations, the determination of $M_B$ is based on type Ia supernova data for redshift, $z<0.15$ with the anchors mentioned above \citep{Riess:2016jrr,Riess:2020fzl}. It is also important to include the higher redshift type Ia supernova observations to determine the value of $M_B$. For this purpose, the Pantheon sample for type Ia supernova observations is useful, where the data have the redshift range up to nearly $2.2$ \citep{Pan-STARRS1:2017jku}.

In the literature, there are some attempts to compute $M_B$ from the cosmological point of view \citep{Camarena:2019rmj,Sapone:2020wwz,Kumar:2021djt,Camarena:2021jlr,Gomez-Valent:2021hda,Cai:2021weh}. These studies are mainly based on the type Ia supernovae data like Pantheon \citep{Pan-STARRS1:2017jku} with other combinations of data sets like CMB observations \citep{Planck:2015fie,Planck:2018vyg}, baryon acoustic oscillations (BAO) observations \citep{eBOSS:2020yzd} etc. Some of these methods like in Refs, \citep{Camarena:2019rmj,Camarena:2021jlr} are not completely independent of astrophysical anchors like stellar parallax \citep{vanLeeuwen:2007xw,2018ApJ...855..136R,Greene:2021shv} and masers \citep{Reid:2019tiq,Pihlstrom:2004vg,Gao:2015tqd}. However, a few other methods like in references \cite{Sapone:2020wwz,Kumar:2021djt,Gomez-Valent:2021hda,Cai:2021weh} depend completely on the cosmological data. These methods are either cosmological model dependent or based on the parametrization of $M_B$. Thus it is worthwhile to consider a model independent and non-parametric approach to estimate $M_B$ from the cosmological data and this estimation should be independent of any astrophysical data or any other data.

The motivation of this work is to compute the bounds on $M_B$ with a complete model independent and parameter-free approach from the cosmological data only. For this purpose, we mainly consider the Pantheon sample for the supernova type Ia observations \citep{Pan-STARRS1:2017jku} and the cosmic chronometer data for the Hubble parameter \citep{Jimenez:2001gg,Pinho:2018unz}, because these data are independent of any fiducial cosmological model. For the methodology, we consider the Gaussian process regression (GPR) analysis \citep{williams1995gaussian,GpRasWil,Seikel_2012,Shafieloo_2012,Hwang:2022hla}.

In recent years, GPR is quite frequently used in cosmology \citep{Velasquez-Toribio:2021ufm,Mukherjee:2020vkx,
Vazirnia:2021xuu,Mukherjee:2020ytg,Haridasu:2018gqm,Zheng:2020tau,Liu:2020pfa,Wang:2019yob,Bernardo:2021cxi,Bonilla:2020wbn,Zhang:2018gjb,Wang:2017jdm,Seikel:2013fda,Mukherjee:2021ggf,Ruiz-Zapatero:2022xbv,
Ruiz-Zapatero:2022zpx,Mehrabi_2022,Zhang_2022,Li:2021onq,Escamilla-Rivera:2021rbe,Bernardo:2021mfs,Keeley:2020aym,Liao:2019qoc,Wang:2016iij,Zhang:2016tto,Nair:2013sna,OColgain:2021pyh}. For example, 
in \citep{Velasquez-Toribio:2021ufm,Mukherjee:2020vkx,Vazirnia:2021xuu,Mukherjee:2020ytg,Haridasu:2018gqm}, the cosmographic parameters like Hubble parameter, deceleration parameter, jerk parameter, etc have been constrained from 
the cosmic chronometers, type Ia supernova, and BAO data using GPR. In \citep{Zheng:2020tau,Liu:2020pfa,Wang:2019yob}, constraining cosmic curvature density parameter has been discussed using the gravitational wave(GW) observations from binary neutron star mergers with the future generation of space-based DECi-hertz Interferometer Gravitational-wave Observatory (DECIGO) and other cosmological observations. In \citep{Bernardo:2021cxi,Bonilla:2020wbn,Zhang:2018gjb,Wang:2017jdm,Seikel:2013fda}, the dark energy equation of state and other dark energy properties have been studied from different cosmological observations. In \citep{Mukherjee:2021ggf}, the interaction of dark energy and dark matter has been constrained by different cosmological observations using GPR.

The CMB \citep{Planck:2018vyg} and the BAO \citep{eBOSS:2020yzd} data, on the other hand,  are dependent on a fiducial cosmological model. As the primary motivation of the present work is a model-independent estimation of $M_B$, we do not include these data sets to start with. However, we will see that as BAO and CMB data sets have significantly smaller error margins (standard deviation), their inclusion in the analysis helps obtain tighter constraints on $M_B$. It is important to note that this difference in the error margins is the only effect of the inclusion of the model dependent data sets, as we will see, the mean value of $M_B$ is hardly affected by the addition of the model-dependent data in the analysis.

This paper is organized as follows. In Sec.~\ref{sec-basic}, we mention basic equations related to the cosmological background dynamics. In Sec.~\ref{sec-data}, we mention some details of the observational data used in our analysis. In Sec.~\ref{sec-methods}, we present our model independent and non-parametric methodology to obtain bounds on the $M_B$ parameter from these observational data. In Sec.~\ref{sec-result}, we present our results and discuss the significance of these results. Finally, in Sec.~\ref{sec-summary}, we summarize the work.

\section{Basics}
\label{sec-basic}

\subsection{Basic cosmological relations}

In our entire analysis, we consider that the Universe is spatially homogeneous and isotropic. We further assume that the Universe is spatially flat too. With these two assumptions, the background geometry can be described by the flat Friedmann-Lema\^itre-Robertson-Walker (FLRW) metric given by $dS^{2} = - dt^{2} + a^{2} (t) dR^2$, where $dS$ is the line element of the space-time, $dR$ is the three-dimensional Euclidean line element, $t$ is the cosmic time and $a$ is the cosmic scale factor. In this scenario, the luminosity distance, $d_L$ is related to the Hubble parameter, $H$ with an integration equation given as

\begin{equation}
d_L(z) = c(1+z) \int_0^z \frac{d\tilde{z}}{ H(\tilde{z}) },
\label{eq:H_to_dL}
\end{equation}

\noindent
where $z$ (also, $\tilde{z}$) is the cosmological redshift given as $1+z = \frac{a_0}{a}$, where $a_0$ is the present value of $a$; $c$ is the speed of light in vacuum.

The observed luminosity distance of a type Ia supernova, located at a particular redshift, is related to the observed apparent peak magnitude ($m$) of the supernovae with a simple equation given as

\begin{equation}
m(z) - M_B = 5 \log_{10}{ \left[ \frac{d_L(z)}{ \text{Mpc} } \right] } + 25,
\label{eq:dL_to_m}
\end{equation}

\noindent
where $M_B$ is the peak absolute magnitude of the same supernova. The above equation is independent of any cosmological model and valid for the only assumption that the Universe is spatially homogeneous and isotropic.

Note that, the above equation has a more generalized version with other parameters involved through the stretch color relation (for details see \citep{Tripp:1997wt}). In the Pantheon compilation of type Ia supernova data, the other parameters are marginalized with the zero centralized value \citep{Pan-STARRS1:2017jku}. Since we are considering the Pantheon compilation data, we are using the above equation only.

\subsection{$H(z)$ from $m(z)$}

We use Eq.~\eqref{eq:dL_to_m}, to get luminosity distance from $m$ and the solution is given as

\begin{equation}
d_L(z) = 10^{\frac{1}{5} \left[ m(z)-25-M_B \right] } \hspace{0.2 cm} \text{Mpc}
\label{eq:dL_from_m}
\end{equation}

\noindent
The above equation can be rewritten as a combination of a redshift independent part and a redshift dependent part. For the redshift independent part, we define a parameter, $\beta$ given as

\begin{equation}
\beta = 10^{-\frac{M_B}{5}} \hspace{0.2 cm} \text{Mpc}.
\label{eq:beta}
\end{equation}

\noindent
For the redshift dependent part, we define a quantity, $d_N$ given as

\begin{equation}
d_N(z) = 10^{\frac{1}{5} \left[ m(z)-25 \right] }.
\label{eq:dN}
\end{equation}

\noindent
With the definitions of $\beta$ and $d_N$, the luminosity distance, $d_L$ can be rewritten as

\begin{equation}
d_L(z) = \beta d_N(z).
\label{eq:dL_wrt_beta_dN}
\end{equation}

\noindent
In the above equation, we can see that $d_L$ is linear in $d_N$ and $d_N$ is independent of the $M_B$ parameter because the $M_B$ parameter is absorbed in the constant parameter, $\beta$.

Not only $d_L$, we also need $d'_L = \frac{dd_L}{dz}$ to find $H$. Throughout this paper, the prime denotes the derivative with respect to the redshift, $z$. To compute $d'_L$, we do the differentiation of Eq.~\eqref{eq:dL_wrt_beta_dN} with respect to $z$ and we get

\begin{equation}
d'_L(z) = \beta d'_N(z),
\label{eq:dLp_from_dNp}
\end{equation}

\noindent
where $d'_N$ is given as (by doing differentiation of Eq.~\eqref{eq:dN} with respect to $z$)

\begin{equation}
d'_N(z) = \alpha m'(z) 10^{\frac{1}{5} \left[ m(z)-25 \right] } = \alpha m'(z) d_N(z),
\label{eq:dNp}
\end{equation}

\noindent
with

\begin{equation}
\alpha = \frac{\log_{\hat{e}}{10}}{5}.
\label{eq:alpha}
\end{equation}

To get the Hubble parameter, we have to differentiate Eq.~\eqref{eq:H_to_dL}. By doing this, we get

\begin{eqnarray}
d'_L(z) &=& c \left[ \frac{1+z}{H(z)} + \int_0^z \frac{d\tilde{z}}{ H(\tilde{z}) } \right] \nonumber\\
&& = \frac{c(1+z)}{H(z)} + \frac{d_L(z)}{1+z}.
\label{eq:derivative_d_L_eqn}
\end{eqnarray}

\noindent
From this equation, we get the Hubble parameter given as

\begin{eqnarray}
H(z) &=& \frac{c(1+z)^2}{(1+z)d'_L(z)-d_L(z)} \noindent\\
&=& \frac{c(1+z)^2}{ \beta \left[ (1+z)d'_N(z)-d_N(z) \right] },
\label{eq:H_from_dL_dLprime}
\end{eqnarray}

\noindent
where in the second equality, we have used Eqs.~\eqref{eq:dL_wrt_beta_dN} and~\eqref{eq:dLp_from_dNp}.

Similar to the case for the luminosity distance, here also, we can separate the parameter independent part (which is redshift dependent) and the parameter dependent part (which is redshift independent). For the parameter independent part, we define a quantity, $G$ given as

\begin{equation}
G(z) = \frac{(1+z)^2}{ (1+z)d'_N(z)-d_N(z) }.
\label{eq:G}
\end{equation}

\noindent
For the parameter dependent part, we define a parameter, $F$ given as

\begin{equation}
F = \frac{c}{\beta} = c \hspace{0.2 cm} 10^{\frac{M_B}{5}} \hspace{0.2 cm} \text{Mpc}^{-1},
\label{eq:F}
\end{equation}

\noindent
where in the second equality we have used the definition of $\beta$ from Eq.~\eqref{eq:beta}. Using the above two definitions, the Hubble parameter can be rewritten as

\begin{equation}
H(z) = F G(z).
\label{eq:H_wrt_F_G}
\end{equation}

\noindent
We can see that $H$ is linear in $G$.

\subsection{Propagation of uncertainty}

Using propagation of uncertainty through Eq.~\eqref{eq:dN}, we compute the uncertainty in $d_N$ (denoted by $\Delta d_N$) given as

\begin{eqnarray}
\Delta d_N &=& \sqrt{\text{Var}(d_N)},
\label{eq:delta_dN} \\
\text{Var}(d_N) &=& \left( \frac{\partial d_N}{\partial m} \right)^2 \text{Var}(m) \nonumber\\
&=& \alpha^2 d_N^2 \text{Var}(m) .
\label{eq:var_dN}
\end{eqnarray}

\noindent
Throughout this paper, we denote the $1\sigma$ uncertainty (or equivalently the standard deviation) of a quantity, $Q$ as $\Delta Q$ and the corresponding variance as Var($Q$), where Var($Q$)=$(\Delta Q)^2$. We also denote the covariance between two quantities, $Q_i$ and $Q_j$ as Cov[$Q_i,Q_j$]. Note that if the two quantities are the same, the covariance is the same as the variance i.e. Cov[$Q,Q$]=Var($Q$).

Similarly, using propagation of uncertainty through Eq.~\eqref{eq:dNp}, we compute the uncertainty in $d'_N$ given as

\begin{eqnarray}
d'_N &=& \sqrt{\text{Var}(d'_N)},
\label{eq:delta_dNprime} \\
\text{Var}(d'_N) &=& \left( \frac{\partial d'_N}{\partial m} \right)^2 \text{Var}(m) + \left( \frac{\partial d'_N}{\partial m'} \right)^2 \text{Var}(m') \nonumber\\
&& +2 \frac{\partial d'_N}{\partial m} \frac{\partial d'_N}{\partial m'} \text{Cov}[m,m'] \nonumber\\
&=& \alpha^4 d_N^2 m'^2 \text{Var}(m) + \alpha^2 d_N^2 \text{Var}(m') \nonumber\\
&& + 2 \alpha^3 d_N^2 m' \text{Cov}[m,m'] .
\label{eq:var_dNprime}
\end{eqnarray}

\noindent
Similarly, using propagation of uncertainty through Eqs.~\eqref{eq:dN} and ~\eqref{eq:dNp}, we also compute the covariance between $d_N$ and $d'_N$ given as

\begin{eqnarray}
\text{Cov}[d_N,d'_N] &=& \frac{\partial d_N}{\partial m} \frac{\partial d'_N}{\partial m} \text{Var}(m) + \frac{\partial d_N}{\partial m'} \frac{\partial d'_N}{\partial m'} \text{Var}(m') \nonumber\\
&& + \left( \frac{\partial d_N}{\partial m} \frac{\partial d'_N}{\partial m'} + \frac{\partial d_N}{\partial m'} \frac{\partial d'_N}{\partial m} \right) \text{Cov}[m,m'] \nonumber\\
&=& \alpha^3 d_N^2 m' \text{Var}(m) + \alpha^2 d_N^2 \text{Cov}[m,m'].
\label{eq:Cov_dN_dNp}
\end{eqnarray}

Next, using propagation of uncertainty through Eq.~\eqref{eq:G}, we compute the uncertainty in $G$ given as

\begin{eqnarray}
\Delta G &=& \sqrt{\text{Var}(G)},
\label{eq:delta_G} \\
\text{Var}(G) &=& \left( \frac{\partial G}{\partial d_N} \right)^2 \text{Var}(d_N)+\left( \frac{\partial G}{\partial d'_N} \right)^2 \text{Var}(d'_N) \nonumber\\
&& + 2 \frac{\partial G}{\partial d_N} \frac{\partial G}{\partial d'_N} \text{Cov} [d_N,d'_N] \nonumber\\
&=& \frac{G^4}{(1+z)^4} \text{Var}(d_N) + \frac{G^4}{(1+z)^2} \text{Var}(d'_N) \nonumber\\
&& - 2 \frac{G^4}{(1+z)^3} \text{Cov} [d_N,d'_N].
\label{eq:variance_G}
\end{eqnarray}

Next, using propagation of uncertainty through Eq.~\eqref{eq:H_wrt_F_G}, we compute the uncertainty in $H$ given as

\begin{equation}
\Delta H = \Big{|} \dfrac{\partial H}{\partial G} \Big{|} \Delta G = |F| \Delta G .
\label{eq:delta_H}
\end{equation}

Similarly, using propagation of uncertainty through Eq.~\eqref{eq:dL_wrt_beta_dN}, the uncertainty in $d_L$ can be computed from $\Delta d_N$ given as

\begin{equation}
\Delta d_L = \Big{|} \dfrac{\partial d_L}{\partial d_N} \Big{|} \Delta d_N = |\beta| \Delta d_N .
\label{eq:delta_dL}
\end{equation}

Note that, in this subsection, we have omitted the argument, $z$ in each quantity for the sake of simplicity to write down the equations. So, we should keep in mind that all the equations in this subsection are valid for each redshift point.

\section{Observational data}
\label{sec-data}

\begin{figure}[tbp]
\centering
\includegraphics[width=0.47\textwidth]{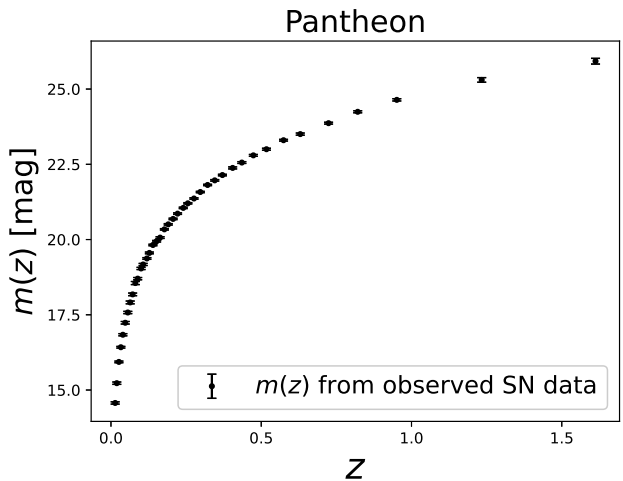}
\caption{
\label{fig:data_SN}
The Pantheon compilation data for the observed peak magnitude, $m(z)$, and the associated uncertainties.
}
\end{figure}

As mentioned in the introduction, in our analysis, we mainly consider two types of observational data. The first one is the Pantheon compilation for the type Ia supernova observations. This compilation consists of data for $m(z)$ at 1048 redshift data points \citep{Pan-STARRS1:2017jku}. Also, these data are binned over 40 redshift bins. We use these binned data in our analysis and denote this as 'SN' data. We are not explicitly writing down all the $m(z)$ values of these data in this paper, because these data are publicly available. To get an idea of the mean values of $m(z)$ and the corresponding uncertainties, see the black error bars in Figure~\ref{fig:data_SN}.

\begin{figure}[tbp]
\centering
\includegraphics[width=0.47\textwidth]{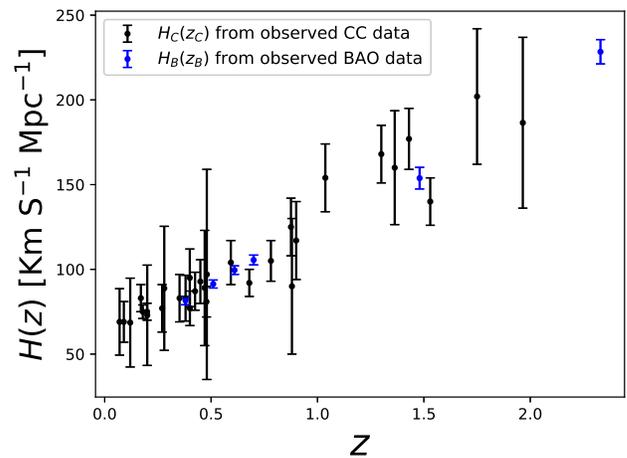}
\caption{
\label{fig:data_H_CC_BAO}
The black colored bars correspond to the CC data for the Hubble parameter and the corresponding uncertainties at CC redshift points. The blue colored bars correspond to the $H_B(z_B)$ and $\Delta H_B(z_B)$ obtained from BAO data.
}
\end{figure}

The second one is the cosmic chronometer data for the Hubble parameter as a function of redshift \citep{Jimenez:2001gg,Pinho:2018unz}. We denote this as 'CC' data and any quantity with subscript 'C' corresponds to the values of that quantity at CC redshift points. These data contain 31 redshift points, the corresponding values of the Hubble parameter, and the corresponding uncertainties. These are plotted in Figure~\ref{fig:data_H_CC_BAO} with black colored bars. The CC data has a redshift range from $0.07$ to $1.965$.

For the sake of completeness, we have also included the BAO data in our analysis. BAO data is not completely model independent because the results have been obtained by considering a fiducial cosmological model. But, it is useful since the error bars in BAO data are smaller compared to the CC data.

The BAO observations consist of measurements for both the line-of-sight direction and the transverse direction \citep{eBOSS:2020yzd}. The line of sight direction data is closely related to the Hubble parameter through the quantity $\tilde{D}_H(z)=D_H(z)/r_d$, where $r_d$ is the comoving sound horizon at the baryon-drag epoch and $D_H(z)=c/H(z)$. The transverse direction data is closely related to the luminosity distance (the comoving angular diameter distance, $D_M(z)$ to be more precise) through the quantity $\tilde{D}_M(z)=D_M(z)/r_d$, where $D_M(z)=d_L(z)/(1+z)$.

So, we need the value of $r_d$, to include BAO data in our analysis. We consider the $r_d$ value obtained from the Planck 2018 result given by $r_d=147.09 \pm 0.26$ Mpc from Planck 2018: TT,TE,EE+lowE+lensing \citep{Planck:2018vyg}, where 'T' stands for temperature in CMB map and 'E' stands for E-modes from CMB polarisation map \citep{Challinor:2012ws,Bucher:2015eia}. The combination of any two quantities corresponds to the power spectrum, for example, 'TE' means the two-point correlation between temperature anisotropy and E-mode polarisation anisotropy \citep{Planck:2018vyg,Challinor:2012ws,Bucher:2015eia}. In this way, we include the CMB data too and we have

\begin{eqnarray}
r_d &=& 147.09 \hspace{0.2 cm} \text{Mpc} , \nonumber\\
\Delta r_d &=& 0.26 \hspace{0.2 cm} \text{Mpc} .
\label{eq:PL18_rd}
\end{eqnarray}

From the above equation, we get $H(z)$ and $\Delta H(z)$ corresponding to the BAO data given as

\begin{eqnarray}
H_{\text{B}}(z_B) &=& \frac{c}{r_d \tilde{D}_H(z_B)}, \nonumber\\
\frac{ \Delta H_B(z_B) }{ H_B(z_B) } &=& \sqrt{ \left[ \frac{\Delta \tilde{D}_H(z_B)}{\tilde{D}_H(z_B)} \right]^2+\left[ \frac{\Delta r_d}{r_d} \right]^2 },
\label{eq:PL18_BAO_H}
\end{eqnarray}

\noindent
respectively. These $H_B(z_B)$ values and the corresponding uncertainties are plotted in Figure~\ref{fig:data_H_CC_BAO} with blue colored bars.

Throughout this paper, subscript 'B' and superscript 'B' to a quantity corresponding to the quantity at the BAO redshift points.

\begin{figure}[tbp]
\centering
\includegraphics[width=0.47\textwidth]{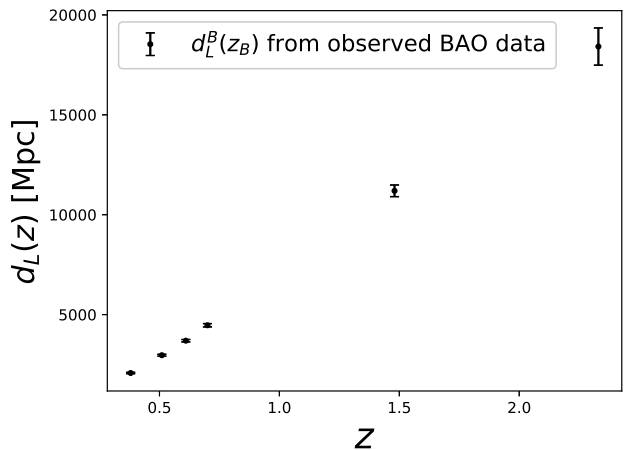}
\caption{
\label{fig:data_dL_BAO}
The $d_L^B(z_B)$ and $\Delta d_L^B(z_B)$ data that obtained from the BAO data.
}
\end{figure}

Similarly, we get $d_L(z)$ and $\Delta d_L(z)$ corresponding to the BAO observations given by

\begin{eqnarray}
d_L^B (z_B) &=& (1+z)r_d \tilde{D}_M(z_B), \nonumber\\
\frac{ \Delta d_L^B (z_B) }{ d_L^B (z_B) } &=& \sqrt{ \left[ \frac{\Delta \tilde{D}_M(z_B)}{\tilde{D}_M(z_B)} \right]^2+\left[ \frac{\Delta r_d}{r_d} \right]^2 },
\label{eq:PL18_BAO_dL}
\end{eqnarray}

\noindent
respectively. The obtained values of $d_L^B(z_B)$ and the corresponding uncertainties are plotted in Figure~\ref{fig:data_dL_BAO} with black colored bars.

Throughout this paper, the BAO data is denoted by the notation 'BAO'. By the 'BAO' notation, we also mean that the value of $r_d$ from Planck 2018 data has been used.

\section{Methodology}
\label{sec-methods}

If we know the observed $m$ and $d_L$ corresponding to a type Ia supernova, we can in principle find its peak absolute magnitude $M_B$ with the help of Eq.~\eqref{eq:dL_to_m}. For this purpose, we have type Ia supernova observations like Pantheon compilation \citep{Pan-STARRS1:2017jku} which provides us the data for $m(z)$. If we consider any theoretical model or any parametrization, we can get a functional form of $d_{L}(z)$ either directly or via the functional form of $H(z)$ through Eq.~\eqref{eq:H_to_dL}. Once we have the functional form of $d_L(z)$, we can put constraints on the parameter $M_B$ (along with other parameters of that model or parametrization). For this case, in principle, it is possible that we can get a constraint on the parameter $M_B$ from only the type Ia supernova observations and this constraint should degenerate to the constraints on other parameters, for example, the Hubble constant, $H_0$ (the Hubble parameter at present i.e. $H_0=H(z=0)$). For better constraints on $M_B$, one can add other data sets.

But, in this analysis, we are not considering any model or any parametrization, rather we want constraints on $M_B$ in a model independent way. Without considering any model or any parametrization, we can not compute $M_B$ with only the type Ia supernova observations. We have to add at least one another type of observation either related to the luminosity distance (or any other quantity closely related to it like the angular diameter distance) or related to the Hubble parameter. For the first case, cosmological observations like BAO \citep{eBOSS:2020yzd} are useful. For the second case, observations like the cosmic chronometers \citep{Jimenez:2001gg,Pinho:2018unz} are useful. Or one can also combine all of these three data.

In general, $d_L(z)$ data (here BAO data) and the $m(z)$ data (here Pantheon compilation) are not at the same redshift points. For this reason, we can not use Eq.~\eqref{eq:dL_to_m} to compute $M_B$ from the combination of these two data sets 
in a straightforward way. For similar reasons, we can not use the Hubble parameter data (here CC data) and the $m(z)$ data together to put a constraint on $M_B$ in a straightforward way.

One possible way to overcome these problems is to use the Gaussian process regression (GPR) technique \citep{williams1995gaussian,GpRasWil,Seikel_2012,Shafieloo_2012,Hwang:2022hla}. This technique is useful to predict the values of any relevant quantity at some target points and the corresponding uncertainties from an observation that consists of data of that quantity at some other points, in general. For example, from $z$, $m$, and $\Delta m$ data points (obtained from the SN data), we can construct values of $m(z)$ and the corresponding uncertainties at CC redshift points.

\subsection{Brief overview of basic GPR analysis}

In GPR, we assume that the observed data, $Y$ (for example, in this case, it is $m$ from SN data) is a multivariate normal distribution, described by only a mean vector and a covariance matrix. The data $Y$ can be expressed by a vector as $Y=[y_1,y_2,...,y_n]^T$, where $y_1,y_2,...,y_n$ are all the observed values at given data points $x_1,x_2,...,x_n$ respectively; $n$ is the number of observed data points. The superscript 'T' represents the transpose of a vector or a matrix. The data points can also be expressed by a vector X given as $X=[x_1,x_2,...,x_n]^T$ (for example, in this case, it is the redshift points of the SN data).

Throughout the discussion, we follow the notation that capital letters correspond to vectors or matrices and the small letters correspond to a single value.

Another important assumption in GPR is that the predicted values of the quantity at some target points also follow a joint multivariate distribution with the data. If the total number of target points is $n^*$, then the joint distribution of data and predicted values has the dimension $n+n*$.

Let us denote the predicted mean vector as $F^*=[f^*_1,f^*_2,...,f^*_{n^*}]^T$ and a covariance matrix as Cov$[F^*,F^*]$, which has $n^* \times n^*$ number of elements. To find these values using GPR, we need an important function, called the kernel covariance function. In literature, there are some forms of this kernel covariance function. Among those, the squared exponential kernel covariance function is the most used. One of the main reasons is that it is infinitely differentiable. In this kernel covariance function, the covariance element corresponding to two points $x_i$ and $x_j$ is expressed as

\begin{equation}
k(x_i,x_j)=\sigma_f^2 \exp \left[ -\frac{|x_i-x_j|^2}{2 l^2} \right],
\label{eq:kernel_SE_main}
\end{equation}

\noindent
where $\sigma_f^2$ is the signal variance that determines the average deviation of a function from its mean along the region of target points and $l$ is the length scale in which the function changes significantly. These parameters are called hyperparameters. In Appendix~\ref{sec-kernels}, we consider other kernel covariance functions and discuss the dependence of the results on these kernels in Appendix~\ref{sec-kernels_dependence}.

We also need prior information for the predictions of GPR through the mean function. In practice, many authors use the zero mean function, but we use the corresponding mean function from the flat $\Lambda$CDM model. In Appendix~\ref{sec-means}, we discuss the form of the mean function for the $\Lambda$CDM model. We also consider other mean functions in Appendix~\ref{sec-means} and show how the results depend on these mean functions in Appendix~\ref{sec-means_dependence}.

Let us denote the values of the mean function at data points and the target points by vectors $M(X)$ (with $n$ number of elements) and $M(X^*)$ (with $n^*$ number of elements) respectively, where $X^*=[x_1^*,x_2^*,...,x_n^*]^T$ is the vector that corresponds to the target points. The predicted mean vector, $F^*$ and the covariance matrix, Cov$[F^*,F^*]$ are given as \citep{Seikel_2012,Shafieloo_2012,Hwang:2022hla}

\begin{eqnarray}
&& F^* = M(X^*) \nonumber\\
&& + K(X^*,X) \left[ K(X,X)+C \right]^{-1} (Y-M(X)), \nonumber\\
&& \text{Cov}[F^*,F^*] = K(X^*,X^*) \nonumber\\
&& - K(X^*,X) \left[ K(X,X)+C \right]^{-1} K(X,X^*),
\label{eq:main_prediction}
\end{eqnarray}

\noindent
respectively. $C$ is the noise covariance matrix of the observed data. Note that the matrix Cov$[F^*,F^*]$ has the elements corresponding to the covariances of all pairs of the elements of $F^*$.

The above equations depend on the values of the hyperparameters of the kernel covariance function and also the parameters of the mean function. We marginalize over all these parameters using the \textit{emcee} package \citep{emcee} with the log marginal likelihood (denoted by $\log P(Y|X)$) given as \citep{Seikel_2012}

\begin{eqnarray}
&& \log P(Y|X) \nonumber\\
&& = -\frac{1}{2} (Y-M(X))^T \left[ K(X,X)+C \right]^{-1} (Y-M(X)) \nonumber\\
&& -\frac{1}{2} \log |K(X,X)+C| -\frac{n}{2} \log{(2 \pi)},
\label{eq:log_marginal_likelihood_main}
\end{eqnarray}

\noindent
where $|K(X,X)+C|$ is the determinant of the $K(X,X)+C$ matrix. The details of the marginalization procedure have been discussed in Appendix~\ref{sec-priors}.

In GPR, the derivatives of the quantity can also be computed by assuming derivatives also follows a joint multivariate normal distribution with the observed data. The mean vector and the covariance matrix corresponding to the first derivative are given as \citep{Seikel_2012}

\begin{eqnarray}
&& F'^* = M'(X^*) \nonumber\\
&& + [K'(X,X^*)]^T \left[ K(X,X)+C \right]^{-1} (Y-M(X)), \nonumber\\
&& \text{Cov}[F'^*,F'^*] = K''(X^*,X^*) \nonumber\\
&& - [K'(X,X^*)]^T \left[ K(X,X)+C \right]^{-1} K'(X,X^*),
\label{eq:derivative_predictions}
\end{eqnarray}

\noindent
where prime and double prime are first and second-order derivatives of the corresponding function respectively with respect to the argument $x$, for example, in our case the redshift. Related to this, $k'(x,x^*)$ and $k''(x^*,x^*)$ are given as

\begin{eqnarray}
k'(x,x^*) = \dfrac{\partial k(x,x^*)}{\partial x^*},
k''(x^*,x^*) = \dfrac{\partial ^2 k(x^*,x^*)}{\partial x^* \partial x^*},
\label{eq:kernel_derivatives_main}
\end{eqnarray}

\noindent
respectively. In GPR, we can also get the covariances between the quantity and its derivatives. For example, the covariance matrix between the quantity and its first derivative is given as \citep{Seikel_2012}

\begin{eqnarray}
&& \text{Cov}[F^*,F'^*] = K'(X^*,X^*) \nonumber\\
&& - [K(X,X^*)]^T \left[ K(X,X)+C \right]^{-1} K'(X,X^*).
\label{eq:covariance_fstar_fstarprime_main}
\end{eqnarray}
\\
\noindent
We have $\text{Cov}[F^*,F'^*] = \left[ \text{Cov}[F'^*,F^*]  \right]^T = \text{Cov}[F'^*,F^*]$, since the covariance matrices are symmetric. More details of the GPR analysis are discussed in Appendix~\ref{sec-GPR_analysis}.

\subsection{Obtaining constraints on $M_B$ from SN and CC data}

We can use GPR to reconstruct $H$ and $\Delta H$ at SN redshift points from the CC data. With the reconstructed $H$ we can reconstruct $m$ as a function of $M_B$ using Eqs.~\eqref{eq:H_to_dL} and~\eqref{eq:dL_to_m}, but the reconstruction of $\Delta m$ is difficult because an integration in Eq.~\eqref{eq:H_to_dL} is not straightforward and there is no standard procedure for ascertaining the propagation of uncertainty through an integration. On the other hand, we can reconstruct $m$ and $\Delta m$ at CC redshift points from SN data using GPR. We can also reconstruct the derivative of $m$ and the corresponding uncertainty using the GPR itself. So we shall choose this method. The details of this method are given below.

\subsubsection{First step: Obtaining $m$ and $m'$ and the corresponding uncertainties at CC redshift points from SN data using GPR}

The SN observations have data of $m$. We denote this as $m(z_S)$ and the corresponding uncertainty as $\Delta m(z_S)$. From here onwards by $z_S$ and $z_C$ we mean the redshift points are at SN and CC data points respectively. In the first step, we use GPR to reconstruct $m(z_C)$, $\Delta m(z_C)$, $m'(z_C)$, $\Delta m'(z_C)$ and Cov[$m(z_C),m'(z_C)$].

\begin{figure}[tbp]
\centering
\includegraphics[width=0.45\textwidth]{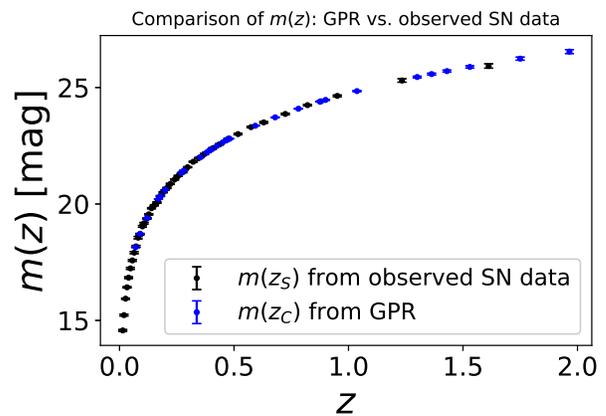}
\caption{
\label{fig:GPR_m_vs_SN_m}
The black colored bars correspond to the Pantheon compilation data for the observed peak magnitude, $m(z_S)$, and the associated errors. The reconstructed $m(z)$ and the corresponding uncertainty at the target CC redshift points by the blue colored bars computed by the GPR analysis.
}
\end{figure}

In Figure~\ref{fig:GPR_m_vs_SN_m}, we have shown the reconstructed mean and uncertainty of $m(z)$ obtained by GPR from the observed SN data. The black colored bars correspond to the SN data from the Pantheon compilation. We have plotted the mean values and the uncertainties of $m$ at target CC redshift points with the blue-colored bars.

\begin{figure}[tbp]
\centering
\includegraphics[width=0.45\textwidth]{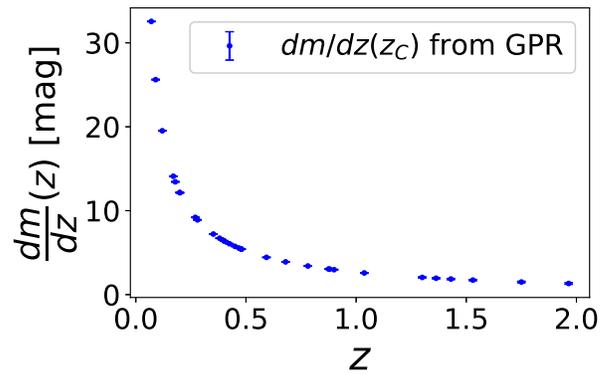}
\caption{
\label{fig:GPR_dmdz_CC}
The reconstructed first derivative of $m$ and the corresponding uncertainties at CC redshift points that obtained using GPR.
}
\end{figure}

In Figure~\ref{fig:GPR_dmdz_CC}, we have plotted the reconstructed $m'=dm/dz$ and the associated uncertainties at CC redshift points obtained using GPR.

\subsubsection{Second step: Obtaining $d_N$ and $d'_N$ and the corresponding uncertainties at CC redshift points}

Here, we compute $d_N$ and $d'_N$ from the reconstructed $m$ and $m'$ (obtained from the previous step) at each CC redshift point using Eqs.~\eqref{eq:dN} and~\eqref{eq:dNp} respectively. Then, we compute the corresponding uncertainties $\Delta d_N$ and $\Delta d'_N$ at each CC redshift point. These are computed from $\Delta m$, $\Delta m'$ and Cov[$m,m'$] (obtained from the previous step) by the propagation of uncertainty using Eqs.~\eqref{eq:delta_dN} and~\eqref{eq:delta_dNprime} respectively. We also compute Cov[$d_N,d'_N$] from $\Delta m$, $\Delta m'$ and Cov[$m,m'$] (obtained from the previous step) by the propagation of uncertainty using Eq.~\eqref{eq:Cov_dN_dNp}.

\begin{figure}[tbp]
\centering
\includegraphics[width=0.45\textwidth]{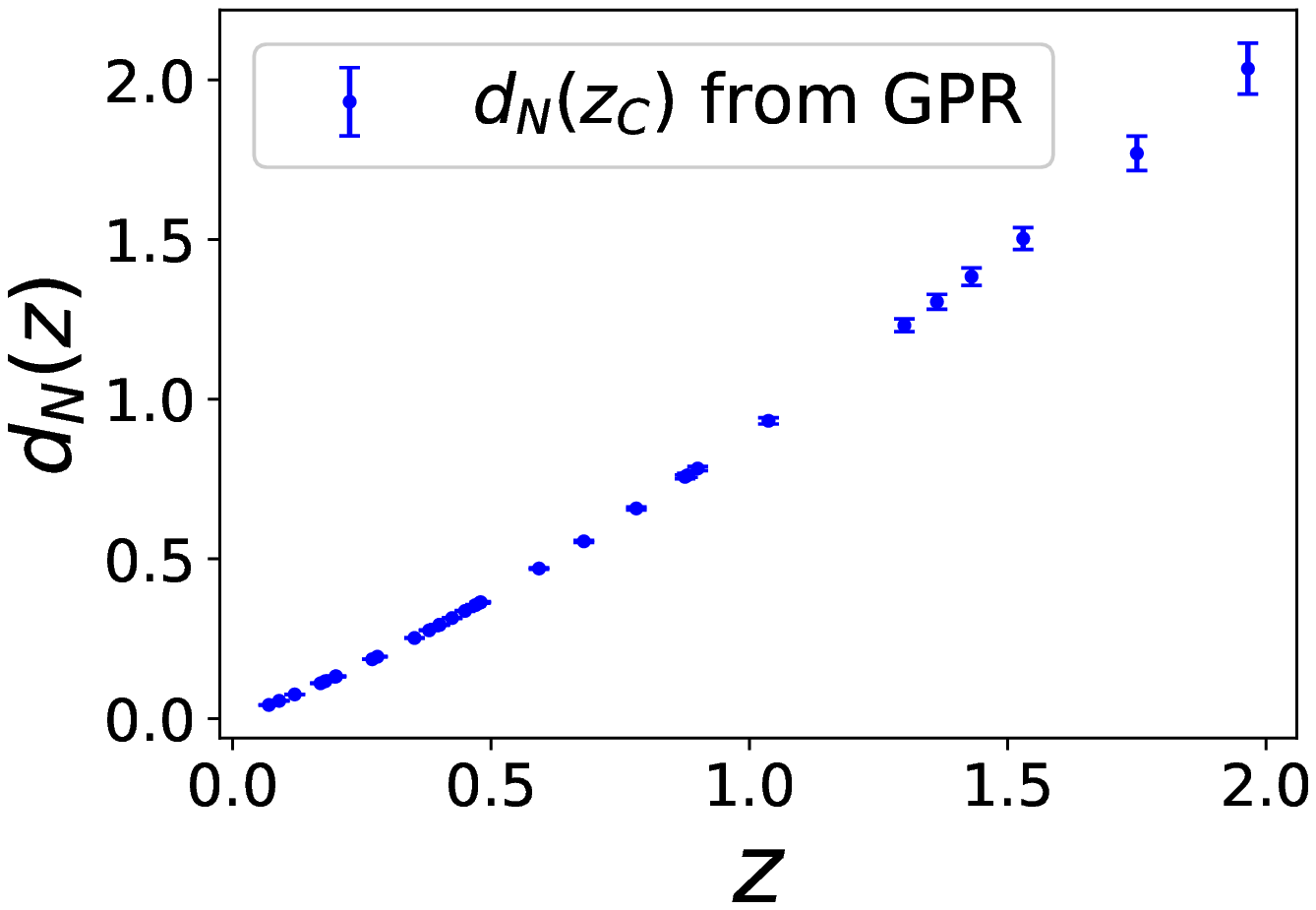}
\caption{
\label{fig:GPR_dN_CC}
Reconstructed $d_N$ and the corresponding uncertainties at CC redshift points.
}
\end{figure}

In Figure~\ref{fig:GPR_dN_CC}, we have plotted the reconstructed $d_N$ and the associated uncertainties at CC redshift points.

\begin{figure}[tbp]
\centering
\includegraphics[width=0.45\textwidth]{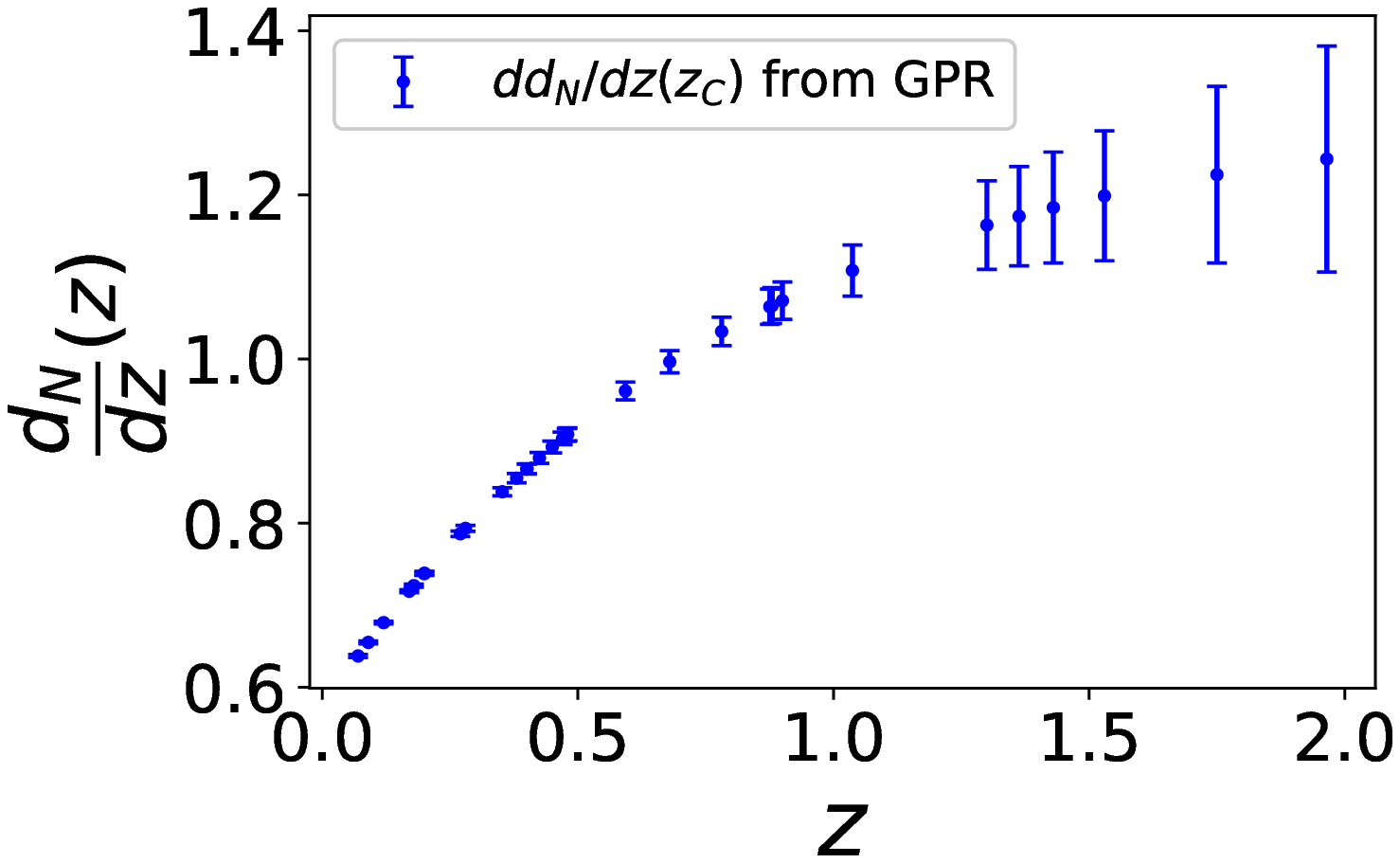}
\caption{
\label{fig:GPR_ddNdz_CC}
Reconstructed $d'_N$ and the corresponding uncertainties at CC redshift points.
}
\end{figure}

In Figure~\ref{fig:GPR_ddNdz_CC}, we have plotted the reconstructed $d'_N$ and the associated uncertainties at CC redshift points.

\subsubsection{Third step: Obtaining $G$ and the corresponding uncertainties at CC redshift points}

Here, we get $G$ from reconstructed $d_N$ and $d'_N$ (obtained from the previous step) at each CC redshift point using Eq.~\eqref{eq:G}. Next, we compute the corresponding uncertainty, $\Delta G$ from $\Delta d_N$, $\Delta d'_N$ and Cov[$d_N,d'_N$] (obtained from the previous step) at each CC redshift point by the propagation of uncertainty using Eq.~\eqref{eq:delta_G}.

\begin{figure}[tbp]
\centering
\includegraphics[width=0.45\textwidth]{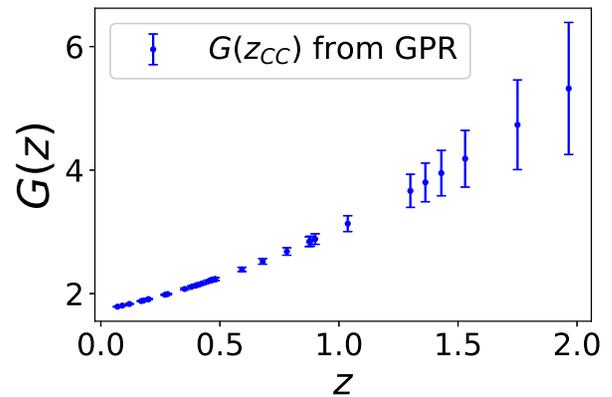}
\caption{
\label{fig:GPR_G}
The mean values and corresponding uncertainties of $G$ at target CC redshift points obtained by GPR are shown by the blue colored bars.
}
\end{figure}

In Figure~\ref{fig:GPR_G}, we have plotted the reconstructed values of $G(z)$ and the corresponding uncertainties at CC redshift points using GPR with the blue colored bars.

\subsubsection{Fourth step: Obtaining constraints on $M_B$ by comparing CC data and the reconstructed $H$ from SN data by GPR}

From the reconstructed $G$ (obtained from the previous step), we get the reconstructed Hubble parameter as a function of $M_B$ parameter (through $F$ parameter) using Eq.~\eqref{eq:H_wrt_F_G} given as

\begin{equation}
H(z_C,M_B)=F(M_B)G(z_C) .
\label{eq:GPR_H_at_CC}
\end{equation}

\noindent
We also get the corresponding uncertainty, $\Delta H$ as a function of $M_B$ from the propagation of uncertainty using Eq.~\eqref{eq:delta_H} given as

\begin{equation}
\Delta H(z_C,M_B) = |F(M_B)| \Delta G(z_C) .
\label{eq:GPR_delH_at_CC}
\end{equation}

Now we compare the reconstructed Hubble parameter to the observed CC data to get constraints on $M_B$. For this purpose, we define a chi-square given as

\begin{equation}
\chi^2_{\text{SN+CC}}(M_B) = \sum_{z_{\text{C}}} \frac{ \left[ H(z_C,M_B)-H_C(z_C) \right]^2}{ \Delta H^2(z_C,M_B) + \Delta H_C^2(z_C) } ,
\label{eq:chisqr_SN_CC}
\end{equation}

\noindent
where $H_C(z_C)$ is the Hubble parameter from the CC data and $\Delta H_C(z_C)$ is the corresponding uncertainty at each CC redshift point. In the above equation, the total term in the denominator corresponds to the total variance in the Hubble parameter. Since the total variance is itself parameter dependent, the better way to get constraints on the parameter is a maximum likelihood analysis rather than the chi-square minimization. The corresponding log-likelihood is given as

\begin{eqnarray}
& \log{L}_{\text{SN+CC}}(M_B) = - \frac{\chi^2_{\text{SN+CC}}(M_B)}{2} \nonumber\\
& - \frac{1}{2} \sum_{z_C} \log{ \left( 2 \pi \left[ \Delta H^2(z_C,M_B) + \Delta H_C^2(z_C) \right] \right) }.
\label{eq:lnlk_SN_CC}
\end{eqnarray}

\noindent
We maximize the likelihood by minimizing the negative log-likelihood to get constraints on $M_B$. In this way, we get constraints on $M_B$ from the combination of SN+CC data.

\begin{figure*}[tbp]
\centering
\begin{tikzpicture}[node distance=2.7cm]
\node(start)[startstop]{SN: $m(z_S)$ and $\Delta m(z_S)$};
\node (in1) [io, below of=start, xshift=-6cm, yshift=0cm] {$m(z_C)$, $\Delta m(z_C)$,
$m'(z_C)$, $\Delta m'(z_C)$,
and Cov[$m(z_C),m'(z_C)$]};
\node (in2) [io, below of=start, xshift=+3cm, yshift=0cm] {$m(z_B)$, $\Delta m(z_B)$,
$m'(z_B)$, $\Delta m'(z_B)$,
and Cov[$m(z_B),m'(z_B)$]};
\node (pro1a) [process, below of=in1, xshift=0cm, yshift=0cm] {$d_N(z_C)$, $\Delta d_N(z_C)$,
$d'_N(z_C)$, $\Delta d'_N(z_C)$,
and Cov[$d_N(z_C),d'_N(z_C)$]};
\node (pro3a) [process, below of=in2, xshift=0cm, yshift=0cm] {$d_N(z_B)$, $\Delta d_N(z_B)$,
$d'_N(z_B)$, $\Delta d'_N(z_B)$,
and Cov[$d_N(z_B),d'_N(z_B)$]};
\node (pro1b) [process2, below of=pro1a, xshift=0cm, yshift=0cm] {$G(z_C)$ and $\Delta G(z_C)$};
\node (pro3b) [process2, below of=pro3a, xshift=-4.5cm, yshift=0cm] {$G(z_B)$ and $\Delta G(z_B)$};
\node (pro4b) [process5, below of=pro3a, xshift=+4.0cm, yshift=0cm] {$d_L(z_B,M_B)$ and $\Delta d_L(z_B,M_B)$};
\node (pro1c) [process3, below of=pro1b, xshift=0cm, yshift=0cm] {$H(z_C,M_B)$ and $\Delta H(z_C,M_B)$};
\node (pro3c) [process3, below of=pro3b, xshift=0cm, yshift=0cm] {$H(z_B,M_B)$ and $\Delta H(z_B,M_B)$};
\node (pro2a) [startstop2, below of=pro1c, xshift=-1.5cm, yshift=0cm] {CC: $H_C(z_C)$ and $\Delta H_C(z_C)$};
\node (pro5a) [startstop3, below of=pro3a, xshift=0cm, yshift=0cm] {BAO: $\tilde{D}_H(z_B)$, $\Delta \tilde{D}_H(z_B)$, $\tilde{D}_M(z_B)$, and $\Delta \tilde{D}_M(z_B)$};
\node (pro4a) [process4, below of=pro5a, xshift=-1.5cm, yshift=0cm] {BAO(H only): $H_B(z_B)$ and $\Delta H_B(z_B)$};
\node (pro6a) [process4, below of=pro5a, xshift=+1.5cm, yshift=0cm] {BAO($d_L$ only): $d_L^B(z_B)$ and $\Delta d_L^B(z_B)$};
\node (dec1) [decision, below of=pro2a, xshift=+1.5cm, yshift=0cm] {SN+CC: $M_B$ and $\Delta M_B$};
\node (dec2) [decision2, below of=pro3c, yshift=0cm] {SN+BAO(H only): $M_B$ and $\Delta M_B$};
\node (dec3) [decision2, below of=pro6a, xshift=+2.0cm, yshift=0cm] {SN+BAO($d_L$ only): $M_B$ and $\Delta M_B$};
\node (dec4) [decision, below of=dec3, xshift=-4cm, yshift=0cm] {SN+BAO: $M_B$ and $\Delta M_B$};
\node (dec5) [decision3, below of=dec4, xshift=-4cm, yshift=0cm] {SN+CC+BAO: $M_B$ and $\Delta M_B$};
\draw [arrow] (start) -- node[anchor=east] {GPR} (in1);
\draw [arrow] (start) -- node[anchor=west] {analysis} (in1);
\draw [arrow] (start) -- node[anchor=east] {GPR} (in2);
\draw [arrow] (start) -- node[anchor=west] {analysis} (in2);
\draw [arrow] (in1) -- node[anchor=east] {Eqs.~\eqref{eq:dN},~\eqref{eq:dNp},} (pro1a);
\draw [arrow] (in1) -- node[anchor=west] {and~\eqref{eq:delta_dN} to~\eqref{eq:Cov_dN_dNp}} (pro1a);
\draw [arrow] (in2) -- node[anchor=east] {Eqs.~\eqref{eq:dN},~\eqref{eq:dNp},} (pro3a);
\draw [arrow] (in2) -- node[anchor=west] {and~\eqref{eq:delta_dN} to~\eqref{eq:Cov_dN_dNp}} (pro3a);
\draw [arrow] (pro1a) -- node[anchor=east] {Eqs.~\eqref{eq:G},} (pro1b);
\draw [arrow] (pro1a) -- node[anchor=west] {~\eqref{eq:delta_G}, and~\eqref{eq:variance_G}} (pro1b);
\draw [arrow] (pro3a) -- node[anchor=east] {Eqs.~\eqref{eq:G},~\eqref{eq:delta_G}, and~\eqref{eq:variance_G}} (pro3b);
\draw [arrow] (pro3a) -- node[anchor=west] {Eqs.~\eqref{eq:GPR_dL_at_BAO} and~\eqref{eq:GPR_del_dL_at_BAO}} (pro4b);
\draw [arrow] (pro1b) -- node[anchor=east] {Eqs.~\eqref{eq:GPR_H_at_CC}} (pro1c);
\draw [arrow] (pro1b) -- node[anchor=west] {and~\eqref{eq:GPR_delH_at_CC}} (pro1c);
\draw [arrow] (pro3b) -- node[anchor=east] {Eqs.~\eqref{eq:GPR_H_at_BAO}} (pro3c);
\draw [arrow] (pro3b) -- node[anchor=west] {and~\eqref{eq:GPR_delH_at_BAO}} (pro3c);
\draw [arrow] (pro1c) -- node[anchor=west] {Eq.~\eqref{eq:lnlk_SN_CC}} (dec1);
\draw [arrow] (pro3c) -- node[anchor=east] {Eq.~\eqref{eq:lnlk_SN_PL18_BAO_H}} (dec2);
\draw [arrow] (pro2a) -- node[anchor=east] {Eq.~\eqref{eq:lnlk_SN_CC}} (dec1);
\draw [arrow] (pro4a) -- node[anchor=west] {Eq.~\eqref{eq:lnlk_SN_PL18_BAO_H}} (dec2);
\draw [arrow] (pro4b) -- node[anchor=west] {Eq.~\eqref{eq:lnlk_SN_PL18_BAO_dL}} (dec3);
\draw [arrow] (pro6a) -- node[anchor=east] {Eq.~\eqref{eq:lnlk_SN_PL18_BAO_dL}} (dec3);
\draw [arrow] (pro5a) -- node[anchor=east] {Eq.~\eqref{eq:PL18_BAO_H}} (pro4a);
\draw [arrow] (pro5a) -- node[anchor=west] {Eq.~\eqref{eq:PL18_BAO_dL}} (pro6a);
\draw [arrow] (dec2) -- node[anchor=west] {Eq.~\eqref{eq:lnlk_SN_PL18_BAO}} (dec4);
\draw [arrow] (dec3) -- node[anchor=east] {Eq.~\eqref{eq:lnlk_SN_PL18_BAO}} (dec4);
\draw [arrow] (dec1) -- node[anchor=west] {Eq.~\eqref{eq:lnlk_SN_CC_PL18_BAO}} (dec5);
\draw [arrow] (dec4) -- node[anchor=east] {Eq.~\eqref{eq:lnlk_SN_CC_PL18_BAO}} (dec5);
\end{tikzpicture}
\caption{
\label{fig:flchrt_SN_CC}
A flowchart to show all the steps of the methodology to obtain constraints on $M_B$ from SN+CC, SN+BAO, and SN+CC+BAO combinations of data.
}
\end{figure*}
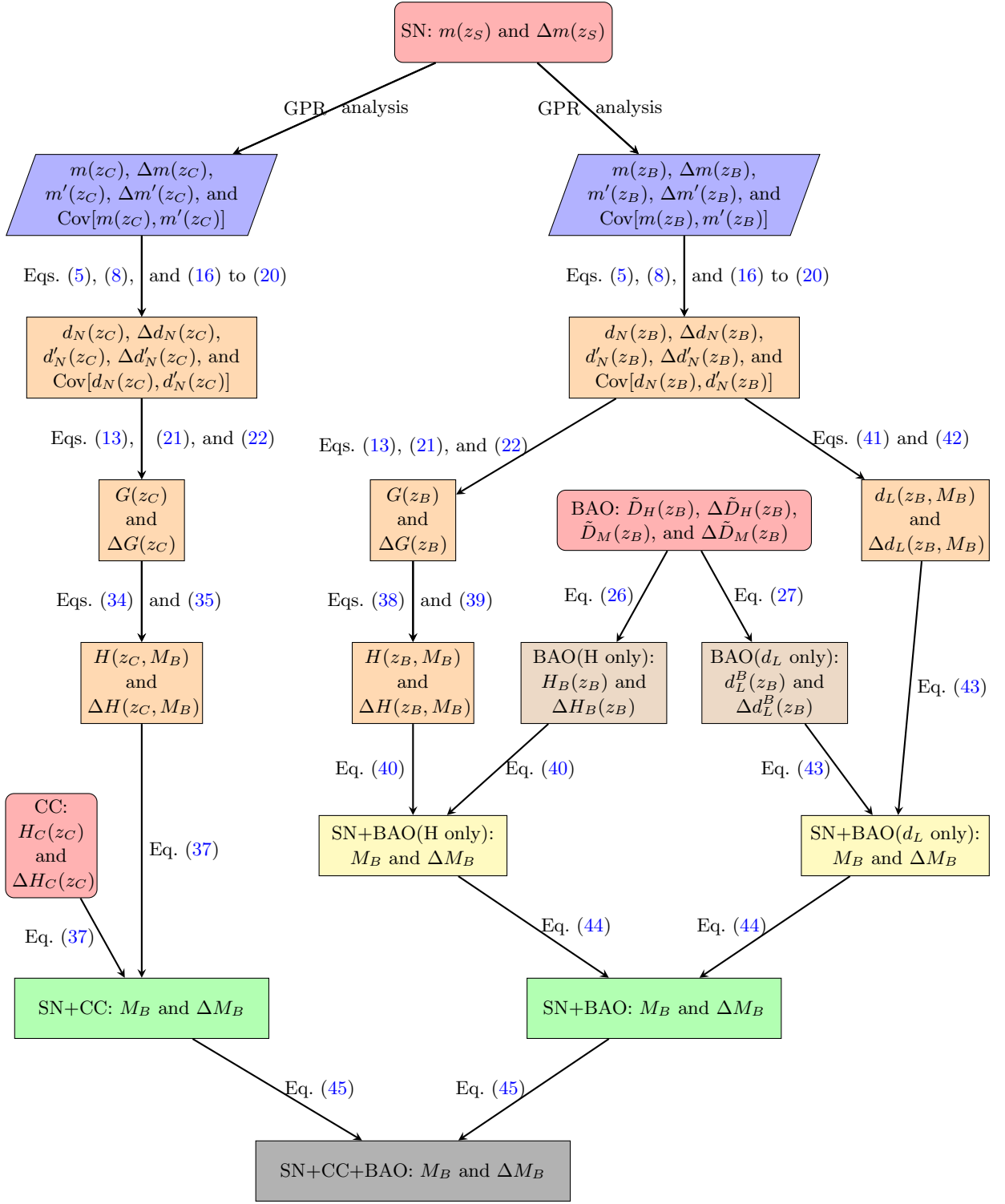

\subsection{Obtaining constraints on $M_B$ from SN and BAO data}
\label{subsec-cmbbao}

In this subsection, we discuss how to include BAO data in our analysis using a similar methodology discussed so far in the previous subsection. As mentioned previously, the BAO observations have two types of data: one is related to the Hubble parameter and the other is related to the luminosity distance. Since we have $H(z)$ data for BAO, we do the same analysis as mentioned in the previous subsection (all the steps from the first step to the fourth step).

First, we get $m(z_B)$, $\Delta m(z_B)$, $m'(z_B)$, $\Delta m'(z_B)$, and Cov[$m(z_B),m'(z_B)$] at each BAO redshift point from SN data using GPR using Eqs.~\eqref{eq:main_prediction},~\eqref{eq:derivative_predictions}, and~\eqref{eq:covariance_fstar_fstarprime_main}.

Next, we get $d_N(z_B)$, $\Delta d_N(z_B)$, $d'_N(z_B)$, $\Delta d'_N(z_B)$, and Cov[$d_N(z_B),d'_N(z_B)$] using Eqs.~\eqref{eq:dN},~\eqref{eq:dNp},~\eqref{eq:delta_dN},~\eqref{eq:delta_dNprime}, and~\eqref{eq:Cov_dN_dNp}.

Then we get $G(z_B)$ and $\Delta G(z_B)$ at each BAO redshift point using Eqs.~\eqref{eq:G} and~\eqref{eq:delta_G}.

From these $G(z_B)$ and $\Delta G(z_B)$, we get the Hubble parameter and the corresponding uncertainty as a function of $M_B$ at each BAO redshift point given as (using Eqs.~\eqref{eq:H_wrt_F_G} and~\eqref{eq:delta_H} respectively)

\begin{eqnarray}
H(z_B,M_B) &=& F(M_B)G(z_B) ,
\label{eq:GPR_H_at_BAO} \\
\Delta H(z_B,M_B) &=& |F(M_B)| \Delta G(z_B) ,
\label{eq:GPR_delH_at_BAO}
\end{eqnarray}

\noindent
respectively. Comparing above equations with Eq.~\eqref{eq:PL18_BAO_H}, we define a corresponding log-likelihood for BAO for $H(z)$ given as

\begin{eqnarray}
& \log{L}_{\text{SN+BAO(H only)}}(M_B) = \nonumber\\
& - \frac{1}{2} \sum_{z_B} \frac{ \left[ H(z_B,M_B)-H_B (z_B) \right]^2}{ \Delta H^2(z_B,M_B) + \Delta H_B^2(z_B) } \nonumber\\
&  - \frac{1}{2} \sum_{z_B} \log{ \left( 2 \pi \left[ \Delta H^2(z_B,M_B) + \Delta H_B^2(z_B) \right] \right) } .
\label{eq:lnlk_SN_PL18_BAO_H}
\end{eqnarray}

Next, from the reconstructed $d_N(z_B)$ and $\Delta d_N(z_B)$, we get the luminosity distance and the corresponding uncertainty at each BAO redshift using Eqs.~\eqref{eq:dL_wrt_beta_dN} and~\eqref{eq:delta_dL} given as

\begin{eqnarray}
d_L(z_B,M_B) &=& \beta(M_B) d_N(z_B) ,
\label{eq:GPR_dL_at_BAO} \\
\Delta d_L(z_B,M_B) &=& |\beta (M_B)| \Delta d_N(z_B) ,
\label{eq:GPR_del_dL_at_BAO}
\end{eqnarray}

\noindent
respectively. Comparing above equations with Eq.~\eqref{eq:PL18_BAO_dL}, we can define a corresponding log-likelihood for BAO $d_L(z)$ given as

\begin{eqnarray}
& \log{L}_{\text{SN+BAO($d_L$ only)}}(M_B) \nonumber\\
& = - \frac{1}{2} \sum_{z_B} \frac{ \left[ d_L(z_B,M_B)-d_L^B (z_B) \right]^2}{ \Delta d_L(z_B,M_B)^2 + (\Delta d_L^B(z_B))^2 } \nonumber\\
& - \frac{1}{2} \sum_{z_{\text{B}}} \log{ \left( 2 \pi \left[ \Delta d_L(z_B,M_B)^2 + (\Delta d_L^B(z_B))^2 \right] \right) } .
\label{eq:lnlk_SN_PL18_BAO_dL}
\end{eqnarray}

Now adding the above two log-likelihoods, we get the total log-likelihood for SN+BAO data given as

\begin{eqnarray}
&& \log{L}_{\text{SN+BAO}}(M_B) = \log{L}_{\text{SN+BAO(H only)}}(M_B) \nonumber\\
&& + \log{L}_{\text{SN+BAO($d_L$ only)}}(M_B).
\label{eq:lnlk_SN_PL18_BAO}
\end{eqnarray}

\noindent
We minimize the negative of the above log-likelihood to get the constraints on $M_B$ from SN+BAO data.

\subsection{Obtaining constraints on $M_B$ from SN, CC and BAO data}

The constraints on $M_B$ from all the data combined i.e. from SN+CC+BAO can be obtained by doing the maximum likelihood analysis for the total log-likelihood given as

\begin{eqnarray}
\log{L}_{\text{SN+CC+BAO}}(M_B) &=& \log{L}_{\text{SN+CC}}(M_B) \nonumber\\
&& + \log{L}_{\text{SN+BAO}}(M_B).
\label{eq:lnlk_SN_CC_PL18_BAO}
\end{eqnarray}

In Figure~\ref{fig:flchrt_SN_CC}, we have shown a flowchart to see all the steps and methods at a glance to obtain constraints on $M_B$ from SN+CC, SN+BAO, and SN+CC+BAO combinations of data.

\section{Results and discussion}
\label{sec-result}

For SN+CC data, we minimize the negative of log-likelihood mentioned in Eq.~\eqref{eq:lnlk_SN_CC} and get constraints on $M_B$ given as

\begin{equation}
M_B = -19.384 \pm 0.052 \hspace{0.2 cm} \text{mag} \hspace{0.2 cm} (\text{SN}+\text{CC}).
\label{eq:result_MB_SN_CC}
\end{equation}

Similarly for SN+BAO data, we minimize the negative log-likelihood mentioned in Eq.~\eqref{eq:lnlk_SN_PL18_BAO} to get constraints on $M_B$ given as

\begin{equation}
M_B = -19.396 \pm 0.016 \hspace{0.2 cm} \text{mag} \hspace{0.2 cm} (\text{SN}+\text{BAO}).
\label{eq:result_MB_SN_BAO_CMB}
\end{equation}

Finally, we get constraints on $M_B$ from all these data combined i.e. from SN+CC+BAO data by minimizing the negative log-likelihood mentioned in Eq.~\eqref{eq:lnlk_SN_CC_PL18_BAO} given as

\begin{equation}
M_B = -19.395 \pm 0.015 \hspace{0.2 cm} \text{mag} \hspace{0.2 cm} (\text{SN}+\text{CC}+\text{BAO}).
\label{eq:result_MB_SN_CC_BAO_CMB}
\end{equation}

\begin{figure}[tbp]
\centering
\includegraphics[width=0.45\textwidth]{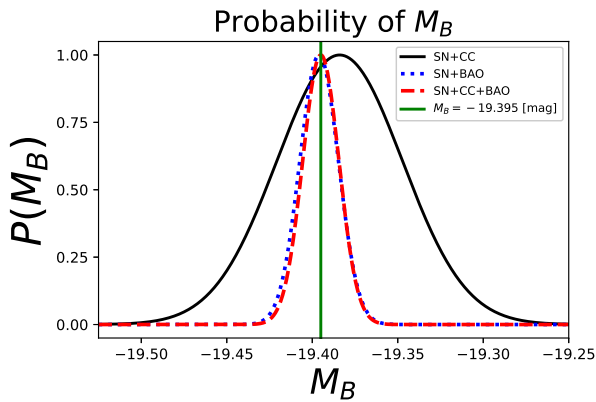}
\caption{
\label{fig:probability_MB}
Probability of $M_B$, obtained from MCMC analysis from log-likelihood accordingly. The solid-black, dotted-blue, and dashed-red lines correspond to the probabilities obtained from SN+CC, SN+BAO, and SN+CC+BAO combinations of data sets respectively. The vertical green line corresponds to the value $M_B = -19.395$ mag.
}
\end{figure}

In Figure~\ref{fig:probability_MB}, we plot the probability of $M_B$ obtained from MCMC analysis from log-likelihood accordingly as mentioned above. The solid-black, dotted-blue, and dashed-red lines correspond to the SN+CC, SN+BAO, and SN+CC+BAO respectively. The vertical green line corresponds to the value $M_B = -19.395$ mag. The constraint on $M_B$ is tighter when we consider SN and BAO data combined compared to the one for SN and CC data combined. This is because the errors on $H(z)$ are significantly smaller in BAO data compared to the CC data. Also in BAO data, the constraints on $M_B$ are coming from the $d_L(z)$ data too which further tightens it. Since, the constraint on $M_B$ is significantly tighter from the BAO data, when we add CC data and BAO data together, the constraints follow the result of BAO data only i.e. there is no significant improvement by adding the CC data. That means for the computation of constraints on $M_B$, if we consider SN and BAO data together, we do not need to add the CC data. But the result from the SN and CC data is important to consider because these data are independent of any fiducial cosmological model whereas the BAO data are dependent on a fiducial model.

There were some efforts to compute $M_B$ from different combinations of different cosmological observations \citep{Camarena:2019rmj,Cai:2021weh,Gomez-Valent:2021hda,Greene:2021shv}. We mention some important previous results below:

In \citep{Camarena:2019rmj}, authors have used a model independent binning technique to combine supernova type Ia observations with the anisotropic BAO observations and find $M_B=-19.401\pm 0.027$ (for details see equation 25 and figure C2 in \citep{Camarena:2019rmj}).

In \citep{Cai:2021weh}, authors have combined supernova type Ia observations, BAO observations, and cosmic chronometer observations to obtain $M_B$. They have used $\Lambda$CDM and PAge \citep{Huang:2020mub} models in their analysis and obtained $M_B=-19.374\pm 0.047$ and $M_B=-19.379^{+0.051}_{-0.052}$ respectively (for details see table I and figure 2 in \citep{Cai:2021weh}).

In \citep{Gomez-Valent:2021hda}, the authors have considered a model independent method to obtain $M_B$ by minimization of a loss function \citep{Lin:2017ikq}. They have combined supernova type Ia observations, BAO observations, and cosmic chronometer observations and obtained $M_B=-19.362^{+0.078}_{-0.067}$ (for details see table I and figure 2 in \citep{Gomez-Valent:2021hda}).

In \citep{Greene:2021shv}, the authors have calibrated type Ia supernova observations with Planck CMB data using $\Lambda$CDM model and obtained $M_B=-19.420\pm0.014$ (for details see figure 7 in \citep{Greene:2021shv}).

We can see the mean values of $M_B$ obtained from all these results are consistent with our results.

\begin{table}
\begin{center}
\begin{tabular}{ |c|c|  }
\hline
Data combinations & Constraints on $M_B$ \\
\hline
SN+CC & $-19.384\pm0.052$ \\
\hline
SN+BAO & $-19.396\pm0.016$ \\
\hline
SN+CC+BAO & $-19.395\pm0.015$ \\
\hline
\end{tabular}
\end{center}
\caption{
Constraints on $M_B$ for different combinations of data.
}
\label{table:MB_main_constraints}
\end{table}

\section{Summary}
\label{sec-summary}

The luminosity of the supernova type Ia is taken as a standard candle in the estimation of cosmic distances in terms of the integrals of the scale factor $a$ and its derivatives. This is crucial in the context of the present state of evolution, particularly the inference regarding the accelerated state of expansion of the universe. This work aims to check the consistency of this assumption by a reconstruction of the peak absolute magnitude, $M_B$, of the type Ia supernova, by a model independent approach from the observational data. Also, the reconstruction is aimed to be independent of any parametrization of cosmological quantities.

We first reconstruct the Hubble parameter at CC redshift points as a function of $M_B$ with the help of the Gaussian process regression (GPR). We also reconstruct the corresponding uncertainties in the Hubble 
parameter, $\Delta_H$ at CC redshift points as another function of $M_B$ using GPR. Note that, in these reconstructions, actual CC data is not involved.

Once we have reconstructed $H$ and $\Delta H(z)$ at CC redshift points, we compare these values with the actual CC data to obtain constraint on $M_B$. We define a corresponding likelihood with the help of Eq.~\eqref{eq:lnlk_SN_CC}. We obtain constraints on $M_B$ by maximizing this likelihood and the result is $M_B = -19.384$ $\pm$ $0.052$ mag.

After this, we deviate from the principal motivation of a model independent work and also include the baryon acoustic oscillation (BAO) data in our analysis. The inclusion of the BAO data makes our analysis model dependent unlike in the case of SN and CC data. Although the mean value of $M_B$ remains quite consistent with the model-independent approach, this inclusion results in tighter constraints on $M_B$.

For the SN+BAO data, we do a similar analysis as in the case of the SN+CC data and we obtain the constraint on $M_B$ as $M_B = -19.396$ $\pm$ $0.016$. Finally, we combine all these three data and get a constraint on $M_B$ as $M_B$ as $M_B = -19.395$ $\pm$ $0.015$. Since SN+BAO data give significantly a tighter constraint compared to the SN+CC data, the result of SN+CC+BAO follows the result of SN+BAO.

We list all these results in Table~\ref{table:MB_main_constraints} for these combinations of data. All the results obtained from different combinations of data mentioned in Eqs.~\eqref{eq:result_MB_SN_CC},~\eqref{eq:result_MB_SN_BAO_CMB}, and~\eqref{eq:result_MB_SN_CC_BAO_CMB} indicate the mean value of $M_B$ to be approximately $-19.4$ (also see Figure~\ref{fig:probability_MB}). These results are similar to the results obtained from previous studies like in \citep{Camarena:2019rmj,Gomez-Valent:2021hda,Cai:2021weh} in the context of similar cosmological data. Note that these results have discrepancies with the results obtained from the astrophysical data like stellar parallax and masers observations like in \citep{Greene:2021shv,Camarena:2021jlr,Dinda:2021ffa,Benisty:2022psx}, in which the results are close to $M_B \approx -19.2$. This discrepancy is already discussed in the literature and it is sometimes referred to as the $M_B$ tension (see \citep{Camarena:2021jlr} for details).

We conclude that the mean value of $M_B$ that is used in the literature is quite consistent with that obtained by the model independent reconstruction. But to obtain tighter constraints, the model dependent tailored data does better. To match that accuracy, we require more data points in the SN and CC data sets.

\appendix

\section{Kernel covariance functions}
\label{sec-kernels}

We will repeat some texts and some equations in the appendix for better flow.

In literature, different kernel covariance functions are used in the GPR analysis. Among those, the squared exponential kernel covariance function is the most used. One of the main reasons is that it is infinitely differentiable. In this kernel covariance function, the covariance between two elements $x_i$ and $x_j$ is expressed as

\begin{equation}
k(x_i,x_j)=\sigma_f^2 \exp \left( -\frac{|x_i-x_j|^2}{2 l^2} \right) ,
\label{eq:kernel_SE}
\end{equation}

\noindent
where $\sigma_f^2$ is the signal variance that determines the average deviation of a function from its mean along the region of target points and $l$ is the length scale in which the function changes significantly. These parameters are called hyper-parameters. We denote this kernel covariance function as 'SE'.

In the main text, we have considered only the squared exponential kernel covariance function. Here, we include some other kernel covariance functions to show how our results depend on different kernel covariance functions. One is the Mat\'ern kernel covariance function with order $5/2$, in which the covariance between two elements $x_i$ and $x_j$ is given as

\begin{equation}
k(x_i,x_j)=\sigma_f^2  \left( 1+\frac{\sqrt{5}d}{l}+\frac{5d^2}{3l^2} \right) \exp \left( -\frac{\sqrt{5}d}{l} \right),
\label{eq:Matern}
\end{equation}

\noindent
where $d=|x_i-x_j|$; $\sigma_f$ and $l$ are two hyper-parameters, similar to the ones for the squared exponential kernel covariance function. We denote this kernel covariance function as 'M5/2'.

Another kernel covariance function is the rational quadratic in which the covariance between two elements $x_i$ and $x_j$ is given as

\begin{equation}
k(x_i,x_j)=\sigma_f^2  \left( 1+\frac{|x_i-x_j|^2}{2 r l^2} \right)^{-r},
\label{eq:rational_quadratic_RQ}
\end{equation}

\noindent
where $\sigma_f$, $l$, and $r$ are three hyper-parameters. All these are non-negative parameters. We denote this kernel covariance function as 'RQ'.

Another kernel covariance function is the periodic in which the covariance between two elements $x_i$ and $x_j$ is given as

\begin{equation}
k(x_i,x_j)=\sigma_f^2 \exp \left[ -\frac{2 \sin^2{ \left( \frac{ \pi |x_i-x_j| }{r} \right) } }{l^2} \right],
\label{eq:periodic}
\end{equation}

\noindent
where $\sigma_f$, $l$, and $r$ are three non-negative hyper-parameters.

\section{Mean functions}
\label{sec-means}

In the flat FLRW metric, the Hubble parameter is given as

\begin{equation}
\frac{H^2 (z) }{H_0^2} = \Omega_{\rm m0}(1+z)^{3}+(1-\Omega_{\rm m0})f_{\rm DE} (z),
\label{eq:DE_Hsqr}
\end{equation}

\noindent
where $\Omega_{\rm m0}$ is the matter-energy density parameter and $f_{\rm DE}$ is given as

\begin{equation}
f_{\rm DE} (z) = \exp \left[ 3 \int_0^z \frac{1+w(\tilde{z})}{1+\tilde{z}} d\tilde{z} \right],
\label{eq:f_DE}
\end{equation}

\noindent
where $w$ is the equation of state of the dark energy. In the main text, we have considered only the $\Lambda$CDM for the dark energy model, where the equation of state of the dark energy is $-1$. Here, we include other three classes of dark energy parametrizations given as the wCDM parametrization, the Chevallier-Polarski-Linder (CPL) parametrization \citep{Chevallier:2000qy,Linder:2002et}, and the Barboza-Alcaniz (BA) parametrization \citep{Barboza:2008rh}. In these parametrizations, the equation of state of the dark energy is given as

\begin{eqnarray}
w (z) \hspace{0.1 cm} (\text{$\Lambda$CDM}) &=& -1, \\
w (z) \hspace{0.1 cm} (\text{wCDM}) &=& w_0, \\
w (z) \hspace{0.1 cm} (\text{CPL}) &=& w_0+w_a \frac{z}{1+z}, \\
w (z) \hspace{0.1 cm} (\text{BA}) &=& w_0+w_a \frac{z(1+z)}{1+z^2},
\label{eq:w_DE_all}
\end{eqnarray}

\noindent
where $w_0$ and $w_a$ are the model parameters.

With the expression of the Hubble parameter in Eq.~\eqref{eq:DE_Hsqr}, we get the luminosity distance, $d_L$ through Eq.~\eqref{eq:H_to_dL} and consequently the apparent magnitude, $m$ of the type Ia supernova through Eq.~\eqref{eq:dL_to_m}. The apparent magnitude, $m$ can be rewritten as

\begin{equation}
m(z) = h_P+5 \log_{10}{ \left[ d_L^{\rm main}(z) \right] },
\label{eq:dL_to_m_rewrite}
\end{equation}

\noindent
where $d_L^{\rm main}$ is given as

\begin{equation}
d_L^{\rm main}(z) = (1+z) \int_0^z \frac{d\tilde{z}}{ H(\tilde{z}) } ,
\end{equation}

\noindent
and $h_P$ is given as

\begin{equation}
h_P = 5 \log_{10}{ \left[ \frac{c}{ H_0 \text{Mpc} } \right] } +25+M_B .
\end{equation}

\noindent
The reason to rewrite $m$ is to show there is degeneracy in $H_0$ and $M_B$ in the expression of $m$ in these parametrizations. So, we have defined a combined parameter, $h_P$ in the above equation.

\section{Gaussian process regression analysis}
\label{sec-GPR_analysis}

\subsection{Basic GPR predictions}

Let us briefly discuss the Gaussian process regression (GPR) analysis. In GPR, we assume that the observed data of a particular function, $f$ (for example, it is $m$ from SN data) is a multivariate normal distribution, described by only a mean vector and a covariance matrix. The mean values of the data are expressed by a vector $Y$ given as $Y=[y_1,y_2,...,y_n]^T$, where $y_1,y_2,...,y_n$ are all the observed values at given data points $x_1,x_2,...,x_n$ respectively (for example, it is the redshift points of the SN data); $n$ is the number of observed data points. The superscript 'T' represents the transpose of a vector or a matrix. The data points are expressed by a vector X given as $X=[x_1,x_2,...,x_n]^T$. So, given the observational data points, the data is assumed to follow the multivariate normal distribution (denoted by $\mathcal{N}$) given as

\begin{equation}
P(Y|X) \sim \mathcal{N} \Big{(} Y \Big{|} M(X),K(X,X) \Big{)},
\label{eq:multi_norm_assm_data}
\end{equation}

\noindent
where $M(X)$ is the mean vector at observational data points corresponding to a chosen mean function, $\mu(x)$ given as

\begin{equation}
M(X) = [\mu(x_1),\mu(x_2),...,\mu(x_n)]^T,
\label{eq:mean_func_at_data}
\end{equation}

\noindent
and $K(X,X)$ is the covariance matrix at observational data points corresponding to a chosen kernel covariance function, $k(x_i,x_j)$ given as

\begin{equation}
K(X,X) = \begin{bmatrix} 
	k(x_1,x_1) & k(x_1,x_2) & ... & k(x_1,x_n) \\
	k(x_2,x_1) & k(x_2,x_2) & ... & k(x_2,x_n) \\
	. & . & ... & . \\
	. & . & ... & . \\
	. & . & ... & . \\
	k(x_n,x_1) & k(x_n,x_2) & ... & k(x_n,x_n) \\
	\end{bmatrix} .
\label{eq:K_data_data}
\end{equation}

\noindent
If observational uncertainty is present, that can be added to the covariance matrix in the distribution of $Y$.

GPR can predict the mean values of the quantity, $f$ at some target points (which are, in general, different from the observational data points), and the values of the associated uncertainty, $\Delta f$. For example, we need predicted mean values of $m$ and the values of associated uncertainty, $\Delta m$ at CC redshift points. Let us consider the target points $x^*_1,x^*_2,...,x^*_{n^*}$ are described by a vector $X^*$ given as $X^*=[x^*_1,x^*_2,...,x^*_{n^*}]^T$, where $n^*$ being the total number of target points. Let us consider the predicted mean vector to be $F^*$ given as $F^*(X^*) = [f^*_1, f^*_2, ..., f^*_{n^*}]^T$ and the associated uncertainties by a matrix, $U^*$ given as

\begin{equation}
U^*(X^*,X^*) = \begin{bmatrix} 
	u^*_{11} & u^*_{12} & ... & u^*_{1n^*} \\
	u^*_{21} & u^*_{22} & ... & u^*_{2n^*} \\
	. & . & ... & . \\
	. & . & ... & . \\
	. & . & ... & . \\
	u^*_{n^*1} & u^*_{n^*2} & ... & u^*_{n^*n^*} \\
	\end{bmatrix} ,
\label{eq:predict_cov_matrix}
\end{equation}

\noindent
where, $u^*_{ij}=\text{Cov}[f^*_i, f^*_j]$ is the covariance between $f^*_i$ and $f^*_j$ ($\forall \hspace{0.1 cm} i,j \in [1,2,...,n^*]$).

GPR predicts $F^*$ and $U^*$ by the assumption that the predicted values also follow a joint multivariate normal distribution with the observed data given as

\begin{align}
\label{eq:joint_pdf_noisy}
& \begin{bmatrix} 
	Y \\
	F^* \\
	\end{bmatrix} \sim \mathcal{N} \Big{(} \begin{bmatrix} 
	M(X) \\
	M(X^*) \\
	\end{bmatrix}, \nonumber\\
	& \begin{bmatrix} 
	K(X,X)+C & K(X,X^*) \\
	K(X^*,X) & K(X^*,X^*) \\
	\end{bmatrix} \Big{)} ,
\end{align}

\noindent
with the number of dimensions to be $n+n^*$. $M(X^*)$ is the vector consisting of the values of the chosen mean function at the target points given as

\begin{equation}
M(X^*) = [\mu(x^*_1),\mu(x^*_2),...,\mu(x^*_{n^*})]^T .
\label{eq:mean_func_at_target}
\end{equation}

\noindent
$C$ is the uncertainty matrix that corresponds to the observational uncertainties given as

\begin{equation}
C = \begin{bmatrix} 
	c_{11} & c_{12} & ... & c_{1n} \\
	c_{21} & c_{22} & ... & c_{2n} \\
	. & . & ... & . \\
	. & . & ... & . \\
	. & . & ... & . \\
	c_{n1} & c_{n2} & ... & c_{nn} \\
	\end{bmatrix} ,
\label{eq:data_noise}
\end{equation}

\noindent
where $c_{ij}=\text{Cov}[y_i,y_j]$ ($\forall \hspace{0.1 cm} i,j \in [1,2,...,n]$). If the uncertainties in the data do not have any correlation between two different data points, the off-diagonal elements in $C$ would be zero. And if there is no uncertainty in the data, all the elements in $C$ would be zero.

Once we choose the mean function and the kernel covariance function, the predicted mean vector, $F^*$, and the uncertainty matrix, $U^*$ are computed as \citep{Seikel_2012,Shafieloo_2012,Hwang:2022hla}

\begin{align}
\label{eq:main_prediction_mean}
F^* &= M(X^*) \nonumber\\
& + K(X^*,X) \left[ K(X,X)+C \right]^{-1} (Y-M(X)), \\
\label{eq:main_prediction_cov}
U^* &= K(X^*,X^*) \nonumber\\
& - K(X^*,X) \left[ K(X,X)+C \right]^{-1} K(X,X^*),
\end{align}

\noindent
respectively. $U^*$ is the same as the Cov[$F^*,F^*$] in the main text in Eq.~\eqref{eq:main_prediction}. In the above equations, the $K(X,X^*)$ matrix is given as

\begin{align}
\label{eq:K_data_star}
& K(X,X^*) = \nonumber\\
& \begin{bmatrix} 
	k(x_1,x^*_1) & k(x_1,x^*_2) & ... & k(x_1,x^*_{n^*}) \\
	k(x_2,x^*_1) & k(x_2,x^*_2) & ... & k(x_2,x^*_{n^*}) \\
	. & . & ... & . \\
	. & . & ... & . \\
	. & . & ... & . \\
	k(x_n,x^*_1) & k(x_n,x^*_2) & ... & k(x_n,x^*_{n^*}) \\
	\end{bmatrix} .
\end{align}

\noindent
The $K(X^*,X)$ matrix is given as

\begin{equation}
K(X^*,X) = [K(X,X^*)]^T .
\label{eq:K_star_data}
\end{equation}

\noindent
Note that the above equation is valid when the chosen kernel covariance function is symmetric over its two arguments i.e. $k(x_i,x_j)=k(x_j,x_i)$. Similarly, $K(X^*,X^*)$ matrix is given as

\begin{align}
\label{eq:K_star_star}
& K(X^*,X^*) = \nonumber\\
& \begin{bmatrix} 
	k(x^*_1,x^*_1) & k(x^*_1,x^*_2) & ... & k(x^*_1,x^*_{n^*}) \\
	k(x^*_2,x^*_1) & k(x^*_2,x^*_2) & ... & k(x^*_2,x^*_{n^*}) \\
	. & . & ... & . \\
	. & . & ... & . \\
	. & . & ... & . \\
	k(x^*_{n^*},x^*_1) & k(x^*_{n^*},x^*_2) & ... & k(x^*_{n^*},x^*_{n^*}) \\
	\end{bmatrix} .
\end{align}

\subsection{Derivative predictions}

In GPR, the derivatives of the function can also be computed by assuming derivatives also follows a joint multivariate normal distribution with the observed data. For example, for the first derivative of the function, $f$, we assume predicted first derivative values follow a joint multivariate normal distribution with the predicted values of the function and with the observed data jointly given as

\begin{align}
\label{eq:joint_pdf_first_derivative}
& \begin{bmatrix} 
	Y \\
	F^* \\
	F'^* \\
	\end{bmatrix} \sim \mathcal{N} \Big{(} \begin{bmatrix} 
	M(X) \\
	M(X^*) \\
	M'(X^*) \\
	\end{bmatrix}, \nonumber\\
	& \begin{bmatrix} 
	K(X,X)+C & K(X,X^*) & K'(X,X^*) \\
	[K(X,X^*)]^T & K(X^*,X^*) & K'(X^*,X^*) \\
	[K'(X,X^*)]^T & K'(X^*,X^*) & K''(X^*,X^*) \\
	\end{bmatrix} \Big{)} ,
\end{align}

\noindent
where primed and double-primed entities are first and second-order derivatives of the corresponding function respectively with respect to any argument, for example, in our case the redshift. In the above equation, the predicted values of the first derivative of the function, $f$ at target points are denoted by a vector $F'^*(X^*)$ given as

\begin{equation}
F'^*(X^*) = [f'^*_1, f'^*_2, ..., f'^*_{n^*}]^T ,
\end{equation}

\noindent
and the values of the derivative of the chosen mean function at target points are denoted by a vector, $M'(X^*)$ given as

\begin{equation}
M'(X^*) = [\mu'(x^*_1),\mu'(x^*_2),...,\mu'(x^*_{n^*})]^T,
\label{eq:mean_func_prime_at_target}
\end{equation}

The predicted mean vector and the covariance matrix corresponding to the first derivative are given as \citep{Seikel_2012}

\begin{align}
\label{eq:derivative_predictions_mean}
& F'^* = M'(X^*) \nonumber\\
& + [K'(X,X^*)]^T \left[ K(X,X)+C \right]^{-1} (Y-M(X)), \\
\label{eq:derivative_predictions_cov}
& V^* = K''(X^*,X^*) \nonumber\\
& - [K'(X,X^*)]^T \left[ K(X,X)+C \right]^{-1} K'(X,X^*),
\end{align}

\noindent
where the covariance matrix, $V^*$ has the structure given as

\begin{equation}
V^*(X^*,X^*) = \begin{bmatrix} 
	v^*_{11} & v^*_{12} & ... & v^*_{1n^*} \\
	v^*_{21} & v^*_{22} & ... & v^*_{2n^*} \\
	. & . & ... & . \\
	. & . & ... & . \\
	. & . & ... & . \\
	v^*_{n^*1} & v^*_{n^*2} & ... & v^*_{n^*n^*} \\
	\end{bmatrix} ,
\label{eq:predict_cov_matrix_first_deriv}
\end{equation}

\noindent
The elements of the $V^*$ matrix represent the covariance of the derivative of the function between two different target points. For example, $v^*_{ij}=\text{Cov}[f'^*_i, f'^*_j]$ corresponds to the covariance of the derivative of $f$ between $x^*_i$ and $x^*_j$ target points. $V^*$ is the same as the Cov[$F'^*,F'^*$] in the main text in Eq.~\eqref{eq:derivative_predictions}. $K'(X,X^*)$ matrix has the structure given as

\begin{align}
\label{eq:Kprime_data_star}
& K'(X,X^*) = \nonumber\\
& \begin{bmatrix} 
	k'(x_1,x^*_1) & k'(x_1,x^*_2) & ... & k'(x_1,x^*_{n^*}) \\
	k'(x_2,x^*_1) & k'(x_2,x^*_2) & ... & k'(x_2,x^*_{n^*}) \\
	. & . & ... & . \\
	. & . & ... & . \\
	. & . & ... & . \\
	k'(x_n,x^*_1) & k'(x_n,x^*_2) & ... & k'(x_n,x^*_{n^*}) \\
	\end{bmatrix} ,
\end{align}

\noindent
$K'(X^*,X^*)$ matrix has the structure given as

\begin{align}
\label{eq:Kprime_star_star}
& K'(X^*,X^*) = \nonumber\\
& \begin{bmatrix} 
	k'(x^*_1,x^*_1) & k'(x^*_1,x^*_2) & ... & k'(x^*_1,x^*_{n^*}) \\
	k'(x^*_2,x^*_1) & k'(x^*_2,x^*_2) & ... & k'(x^*_2,x^*_{n^*}) \\
	. & . & ... & . \\
	. & . & ... & . \\
	. & . & ... & . \\
	k'(x^*_{n^*},x^*_1) & k'(x^*_{n^*},x^*_2) & ... & k'(x^*_{n^*},x^*_{n^*}) \\
	\end{bmatrix} ,
\end{align}

\noindent
and $K''(X^*,X^*)$ matrix has the structure given as

\begin{align}
\label{eq:K_double_prime_star_star}
& K''(X^*,X^*) = \nonumber\\
& \begin{bmatrix} 
	k''(x^*_1,x^*_1) & k''(x^*_1,x^*_2) & ... & k''(x^*_1,x^*_{n^*}) \\
	k''(x^*_2,x^*_1) & k''(x^*_2,x^*_2) & ... & k''(x^*_2,x^*_{n^*}) \\
	. & . & ... & . \\
	. & . & ... & . \\
	. & . & ... & . \\
	k''(x^*_{n^*},x^*_1) & k''(x^*_{n^*},x^*_2) & ... & k''(x^*_{n^*},x^*_{n^*}) \\
	\end{bmatrix} .
\end{align}

In all the above equations, $k'(x_i,x^*_j)$, $k'(x^*_i,x^*_j)$, and $k''(x^*_i,x^*_j)$ are given as

\begin{eqnarray}
k'(x_i,x^*_j) &=& \dfrac{\partial k(x_i,x^*_j)}{\partial x^*_j}, \\
k'(x^*_i,x^*_j) &=& \dfrac{\partial k(x^*_i,x^*_j)}{\partial x^*_j}, \\
k''(x^*_i,x^*_j) &=& \dfrac{\partial ^2 k(x^*_i,x^*_j)}{\partial x^*_i \partial x^*_j},
\label{eq:kernel_derivatives}
\end{eqnarray}

\noindent
respectively. In GPR, we can also get the covariances between the function and its derivatives. The covariance matrix between the function and its first derivative is given as \citep{Seikel_2012}

\begin{align}
\label{eq:covariance_fstar_fstarprime}
& W^* = K'(X^*,X^*) \nonumber\\
& - [K(X,X^*)]^T \left[ K(X,X)+C \right]^{-1} K'(X,X^*),
\end{align}

\noindent
where $W^*$ matrix has the structure given as

\begin{equation}
W^*(X^*,X^*) = \begin{bmatrix} 
	w^*_{11} & w^*_{12} & ... & w^*_{1n^*} \\
	w^*_{21} & w^*_{22} & ... & w^*_{2n^*} \\
	. & . & ... & . \\
	. & . & ... & . \\
	. & . & ... & . \\
	w^*_{n^*1} & w^*_{n^*2} & ... & w^*_{n^*n^*} \\
	\end{bmatrix} ,
\label{eq:cov_main_vs_first_deriv}
\end{equation}

\noindent
with $w^*_{ij}$ is the covariance between the function at $i$th target point and the first derivative of the function at $j$th target point given as $w^*_{ij}=\text{Cov}[f^*_i, f'^*_j]$. $W^*$ is the same as the Cov[$F^*,F'^*$] in the main text in Eq.~\eqref{eq:covariance_fstar_fstarprime_main}.

\begin{figure*}[tbp]
\centering
\includegraphics[width=0.9\textwidth]{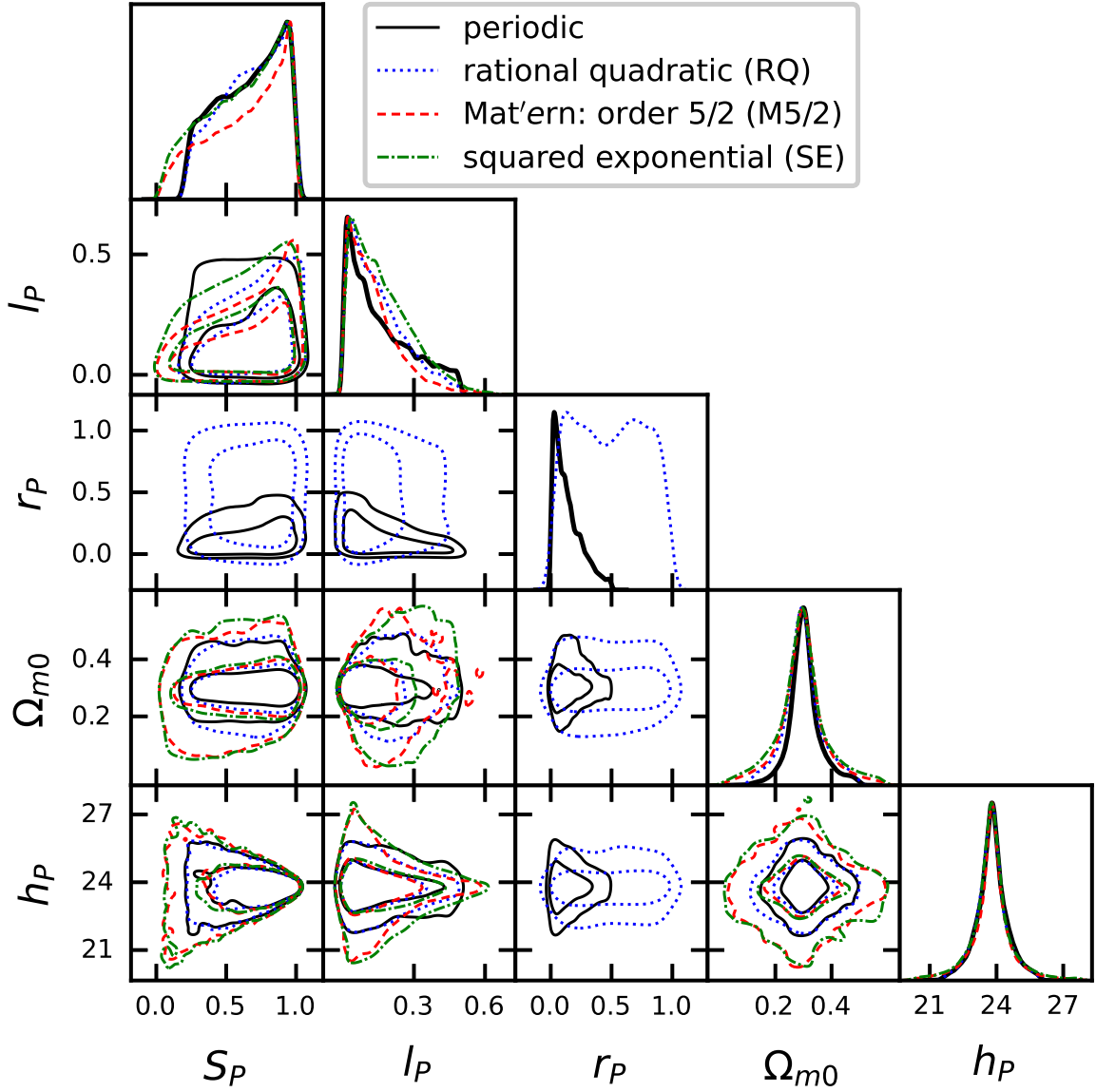}
\caption{
\label{fig:triangle_plot_kernels}
Triangle plot to show the marginalized probability of each parameter and the confidence contours of each pair of parameters for different kernel covariance functions with the $\Lambda$CDM mean function.
}
\end{figure*}

\begin{figure*}[tbp]
\centering
\includegraphics[width=0.9\textwidth]{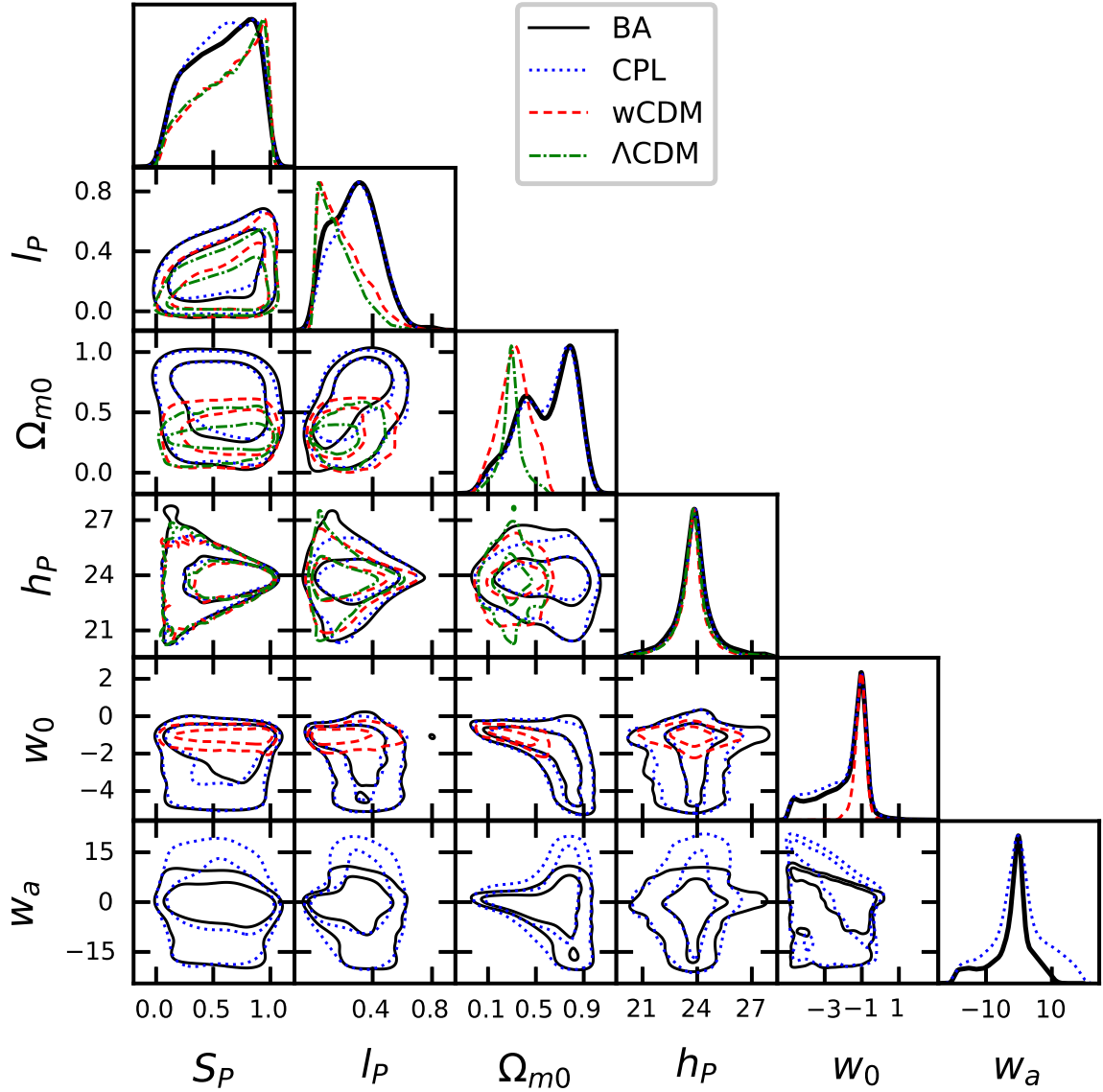}
\caption{
\label{fig:triangle_plot_means}
Triangle plot to show the marginalized probability of each parameter and the confidence contours of each pair of parameters for different mean functions with the squared exponential (SE) kernel covariance function.
}
\end{figure*}

\begin{table}
\begin{center}
\begin{tabular}{ |c|c| }
\hline
Parameters & Priors \\
\hline
$\sigma_f$ & $10^{-10} \leq S_P = \frac{1}{1+\sigma_f^2} \leq 0.99999$ \\
\hline
$l$ & $10^{-10} \leq l_P = \frac{1}{1+l} \leq 0.99999$ \\
\hline
$r$ & $10^{-10} \leq r_P = \frac{1}{1+r} \leq 0.99999$ \\
\hline
$\Omega_{\rm m0}$ & $0.001 \leq \Omega_{\rm m0} \leq 0.999$ \\
\hline
$H_0$ and $M_B$ & $20 \leq h_P \leq 28$ \\
\hline
$w_0$ & $-5 \leq w_0 \leq 3$ \\
\hline
$w_a$ & $-20 \leq w_a \leq 20$ \\
\hline
\end{tabular}
\end{center}
\caption{
Priors on kernel covariance function hyperparameters and mean function parameters.
}
\label{table:priors}
\end{table}

\section{Determination of kernel covariance function hyperparameter and mean function parameter values}
\label{sec-priors}

The GPR predictions through Eqs.~\eqref{eq:main_prediction_mean},~\eqref{eq:main_prediction_cov},~\eqref{eq:derivative_predictions_mean},~\eqref{eq:derivative_predictions_cov}, and~\eqref{eq:covariance_fstar_fstarprime} depend on the values of the hyper-parameters of the chosen kernel covariance function and also the parameters of the chosen mean function. So, we have to put the particular values of these parameters in the above equations to get the predictions of GPR for the mean values and covariances. We can not put the arbitrary values of these parameters. To find these parameter values, we use the knowledge of the observed data which means the parameter values should be chosen such that the values of the chosen mean function at the observed data points closely follow the mean values of the observed data and any differences should be minimum. In practice, this is done by defining a corresponding log marginal likelihood (denoted by $\log P(Y|X)$) given as \citep{Seikel_2012}

\begin{align}
\label{eq:log_marginal_likelihood}
& \log P(Y|X) = \nonumber\\
& -\frac{1}{2} (Y-M(X))^T \left[ K(X,X)+C \right]^{-1} (Y-M(X)) \nonumber\\
& -\frac{1}{2} \log |K(X,X)+C| -\frac{n}{2} \log{(2 \pi)},
\end{align}

\noindent
where $|K(X,X)+C|$ is the determinant of the $K(X,X)+C$ matrix. We minimize the negative log marginal likelihood and find the best-fit values of the parameters. These best-fit values are used to determine the predictions of the GPR. We do this minimization by the Bayesian Markov chain Monte Carlo (MCMC) analysis using the \textit{emcee} package \citep{emcee}, corresponding to the log marginal likelihood, mentioned in Eq.~\eqref{eq:log_marginal_likelihood}. For this purpose, we chose the flat priors on all the parameters according to a chosen kernel covariance function and mean function. We list all the priors in Table~\ref{table:priors}.

The Bayesian MCMC analysis not only gives the best-fit values of the parameters but also the uncertainties around the mean values and the correlation among all the parameters. Let us see these best-fit values and correlations of the parameters through the triangle plots in Figures~\ref{fig:triangle_plot_kernels} and~\ref{fig:triangle_plot_means}.

In Figure~\ref{fig:triangle_plot_kernels}, we have plotted the triangle plot to show the marginalized probability of each parameter and the confidence contours for each pair of the parameters. For a particular color or a particular type of lines, the inner and the outer lines correspond to the 1$\sigma$ and 2$\sigma$ confidence contours respectively. In this figure, we have fixed the mean function to be the $\Lambda$CDM and chosen four different types of kernel covariance functions. The dashed-dotted green, dashed red, dotted blue, and solid black lines correspond to the squared exponential (SE), Mat\'ern with order $5/2$ (M5/2), rational quadratic (RQ), and periodic kernel covariance functions respectively. Corresponding to this, in Table~\ref{table:param_values_kernel}, we list the best-fit values and the 1$\sigma$ marginalized confidence intervals of each parameter.

\begin{table}
\begin{center}
\begin{tabular}{ |c|c|c|c|c|  }
\hline
 & SE & M5/2 & RQ & periodic \\
\hline
$S_P$ & $0.62^{+0.37}_{-0.16}$ & $0.64^{+0.35}_{-0.14}$ & $0.68^{+0.31}_{-0.14}$ & $0.67^{+0.32}_{-0.19}$\\
\hline
$l_P$ & $0.168^{+0.054}_{-0.16}$ & $0.131^{+0.031}_{-0.13}$ & $0.147^{+0.047}_{-0.14}$ & $0.161^{+0.053}_{-0.16}$\\
\hline
$r_P$ & $-$ & $-$ & $0.49\pm 0.29$ & $0.139^{+0.040}_{-0.14}$\\
\hline
$\Omega_{\rm m0}$ & $0.295\pm 0.095$ & $0.296\pm 0.086$ & $0.298^{+0.054}_{-0.065}$ & $0.305^{+0.038}_{-0.053}$\\
\hline
$h_P$ & $23.76\pm 0.99$ & $23.77\pm 0.96$ & $23.76\pm 0.70$ & $23.79\pm 0.72$\\
\hline
\end{tabular}
\end{center}
\caption{
List of best-fit values and the 1$\sigma$ marginalized confidence intervals of each parameter for squared exponential (SE), Mat\'ern with order $5/2$ (M5/2), rational quadratic (RQ), and periodic kernel covariance functions with $\Lambda$CDM mean function.
}
\label{table:param_values_kernel}
\end{table}

In Figure~\ref{fig:triangle_plot_means}, we have plotted the triangle plot to show the marginalized probability of each parameter and the confidence contours for each pair of the parameters. For a particular color or a particular type of lines, the inner and the outer lines correspond to the 1$\sigma$ and 2$\sigma$ confidence contours respectively. In this figure, we have fixed the kernel covariance function to be the squared exponential (SE) and have chosen four different types of mean functions. The dashed-dotted green, dashed red, dotted blue, and solid black lines correspond to the $\Lambda$CDM, wCDM, CPL, and BA mean functions respectively. Corresponding to this, in Table~\ref{table:param_values_mean}, we list the best-fit values and the 1$\sigma$ marginalized confidence intervals of each parameter.

\begin{table}
\begin{center}
\begin{tabular}{ |c|c|c|c|c|  }
\hline
 & $\Lambda$CDM & wCDM & CPL & BA \\
\hline
$S_P$ & $0.62^{+0.37}_{-0.16}$ & $0.63^{+0.37}_{-0.15}$ & $0.57^{+0.36}_{-0.20}$ & $0.57^{+0.36}_{-0.20}$\\
\hline
$l_P$ & $0.168^{+0.054}_{-0.16}$ & $0.200^{+0.069}_{-0.19}$ & $0.29\pm 0.15$ & $0.28^{+0.16}_{-0.18}$\\
\hline
$\Omega_{\rm m0}$ & $0.295\pm 0.095$ & $0.33\pm 0.13$ & $0.59^{+0.29}_{-0.22}$ & $0.59^{+0.31}_{-0.25}$\\
\hline
$h_P$ & $23.76\pm 0.99$ & $23.70\pm 0.91$ & $23.74^{+0.80}_{-0.67}$ & $23.8\pm 1.1$\\
\hline
$w_0$ & $-$ & $-1.08^{+0.31}_{-0.24}$ & $-1.98^{+1.4}_{-0.74}$ & $-1.89^{+1.3}_{-0.64}$\\
\hline
$w_a$ & $-$ & $-$ & $-0.1\pm 8.1$ & $-2.3^{+6.9}_{-3.2}$\\
\hline
\end{tabular}
\end{center}
\caption{
List of best-fit values and the 1$\sigma$ marginalized confidence intervals of each parameter for $\Lambda$CDM, wCDM, CPL, and BA mean functions with squared exponential (SE) kernel covariance function.
}
\label{table:param_values_mean}
\end{table}

In some cases, the predicted variances from the GPR analysis through Eqs.~\eqref{eq:main_prediction_cov} and~\eqref{eq:derivative_predictions_cov} are underestimating. Because of this reason, it is the best practice to include the uncertainties (obtained from the MCMC analysis) in the hyperparameters of the kernel covariance functions and the parameters of the mean functions instead of only considering their best-fit values. So, we compute the propagation of uncertainties in the mean values predictions of GPR through Eqs.~\eqref{eq:main_prediction_mean} and~\eqref{eq:derivative_predictions_mean} from the uncertainties of all the parameters, involved in these equations \citep{Hwang:2022hla}. We do this propagation of uncertainties using the \textit{getdist} package \citep{Lewis:2019xzd}. We add these propagated uncertainties with the GPR predicted uncertainties in Eqs.~\eqref{eq:main_prediction_cov},~\eqref{eq:derivative_predictions_cov}, and~\eqref{eq:covariance_fstar_fstarprime} to find the total covariances and correspondingly the total variances.

\begin{table*}
\begin{tabular}{ c c c c c|c|c|c|c| }
\hline
& SN+CC & & \\
\hline
kernel & $M_B$ & $\% M_B$ & $\% \Delta M_B$ \\
\hline
SE & $-19.384\pm0.052$ & $0.0$ & $0.0$ \\
\hline
M5/2 & $-19.384\pm0.053$ & $0.0$ & $1.9$ \\
\hline
RQ & $-19.385\pm0.053$ & $0.005$ & $1.9$ \\
\hline
Periodic & $-19.392\pm0.053$ & $0.04$ & $1.9$ \\
\hline
\end{tabular}
\begin{tabular}{ c c c c|c|c|c| }
\hline
& SN+BAO & \\
\hline
$M_B$ & $\% M_B$ & $\% \Delta M_B$ \\
\hline
$-19.396\pm0.016$ & $0.0$ & $0.0$ \\
\hline
$-19.396\pm0.016$ & $0.0$ & $0.0$ \\
\hline
$-19.396\pm0.016$ & $0.0$ & $0.0$ \\
\hline
$-19.396\pm0.016$ & $0.0$ & $0.0$ \\
\hline
\end{tabular}
\begin{tabular}{ c c c c|c|c|c| }
\hline
& SN+CC+BAO & \\
\hline
$M_B$ & $\% M_B$ & $\% \Delta M_B$ \\
\hline
$-19.395\pm0.015$ & $0.0$ & $0.0$ \\
\hline
$-19.395\pm0.015$ & $0.0$ & $0.0$ \\
\hline
$-19.395\pm0.015$ & $0.0$ & $0.0$ \\
\hline
$-19.395\pm0.016$ & $0.0$ & $6.7$ \\
\hline
\end{tabular}
\caption{
Values of $M_B$ and $\Delta M_B$ and their percentage deviations for different kernel covariance functions from the corresponding ones for the squared exponential (SE) kernel covariance function. Here, the mean function is fixed to be the $\Lambda$CDM.
}
\label{table:MB_dfrnt_kernels}
\end{table*}

\begin{table*}
\begin{tabular}{ c c c c c|c|c|c|c| }
\hline
& SN+CC & & \\
\hline
mean & $M_B$ & $\% M_B$ & $\% \Delta M_B$ \\
\hline
$\Lambda$CDM & $-19.384\pm0.052$ & $0.0$ & $0.0$ \\
\hline
wCDM & $-19.390\pm0.053$ & $0.03$ & $1.9$ \\
\hline
CPL & $-19.395\pm0.054$ & $0.06$ & $3.8$ \\
\hline
BA & $-19.395\pm0.055$ & $0.06$ & $5.8$ \\
\hline
\end{tabular}
\hspace{0.0cm}
\begin{tabular}{ c c c c|c|c|c| }
\hline
& SN+BAO & \\
\hline
$M_B$ & $\% M_B$ & $\% \Delta M_B$ \\
\hline
$-19.396\pm0.016$ & $0.0$ & $0.0$ \\
\hline
$-19.400\pm0.016$ & $0.02$ & $0.0$ \\
\hline
$-19.406\pm0.017$ & $0.05$ & $6.3$ \\
\hline
$-19.405\pm0.016$ & $0.05$ & $0.0$ \\
\hline
\end{tabular}
\hspace{0.0cm}
\begin{tabular}{ c c c c|c|c|c| }
\hline
& SN+CC+BAO & \\
\hline
$M_B$ & $\% M_B$ & $\% \Delta M_B$ \\
\hline
$-19.395\pm0.015$ & $0.0$ & $0.0$ \\
\hline
$-19.398\pm0.016$ & $0.02$ & $6.7$ \\
\hline
$-19.405\pm0.016$ & $0.05$ & $6.7$ \\
\hline
$-19.404\pm0.016$ & $0.05$ & $6.7$ \\
\hline
\end{tabular}
\caption{
Values of $M_B$ and $\Delta M_B$ and their percentage deviations for different mean functions from the corresponding ones for the $\Lambda$CDM mean function. Here, the kernel covariance function is the squared exponential (SE).
}
\label{table:MB_dfrnt_means}
\end{table*}

\section{Dependence of GPR predictions on kernel covariance functions}
\label{sec-kernels_dependence}

Here we show how much the results change if we choose different kernel covariance functions. For this purpose, we list the mean values of $M_B$ and the corresponding uncertainties in Table~\ref{table:MB_dfrnt_kernels} obtained from SN+CC, SN+BAO, and SN+CC+BAO combinations of data for different kernel covariance functions with $\Lambda$CDM mean function. We see the results are very similar. We have also shown the percentage deviations. The notation, $\% M_B$ corresponds to

\begin{equation}
\% M_B = \left[ \frac{M_B}{M_B(\text{SE: $\Lambda$CDM})} -1 \right] \times 100. \nonumber
\end{equation}

\noindent
The notation, $\% \Delta M_B$ corresponds to

\begin{equation}
\% \Delta M_B = \left[ \frac{\Delta M_B}{\Delta M_B(\text{SE: $\Lambda$CDM})} -1 \right] \times 100. \nonumber
\end{equation}

From Table~\ref{table:MB_dfrnt_kernels}, we can see that mean values of $M_B$ differ at sub-percentage levels for different kernel covariance functions. The uncertainties in $M_B$ differ within $10\%$ for different kernel covariance functions.

\section{Dependence of GPR predictions on mean functions}
\label{sec-means_dependence}

Here, we show how much the results change if we choose different mean functions. For this purpose, we list the percentage deviations in $M_B$ and $\Delta M_B$, as in the previous subsection, in Table~\ref{table:MB_dfrnt_means} for SN+CC, SN+BAO, and SN+CC+BAO combinations of data for four different mean functions, mentioned in these tables. Here we have fixed the kernel covariance function to be the squared exponential (SE). We see that the deviations in the mean values of $M_B$ are within $0.1\%$ and the deviations in the uncertainties are within $10\%$.

\section{Full Pantheon data versus the binned data}

So far, we have used the binned version of the Pantheon compilation for the type Ia supernova observations in our entire analysis. The results would be similar if we consider the full pantheon sample. Because, in the construction of the binned version of the data from the full pantheon sample, the errors are considered accordingly for the redshift points binning. To show this fact, now, we have considered the full pantheon sample and followed the same analysis as in the main text to find constraints on $M_B$. We have listed these values in Table~\ref{table:FuLL_Panth_vs_bin} for SN+CC, SN+BAO, SN+CC+BAO combinations of data with the squared exponential kernel covariance function and the $\Lambda$CDM mean function. We have also shown the percentage deviations in the results compared to the results obtained from the binned data. We can see that the results are very similar.

\begin{table}
\begin{center}
\begin{tabular}{ |c|c|c|c|c|  }
\hline
 & $M_B$ (full) & $\% M_B$ & $\% \Delta M_B$ \\
\hline
SN+CC & $-19.379\pm0.052$ & $-0.03$ & $0.0$ \\
\hline
SN+BAO & $-19.391\pm0.016$ & $-0.03$ & $0.0$ \\
\hline
SN+CC+BAO & $-19.390\pm0.015$ & $-0.03$ & $0.0$ \\
\hline
\end{tabular}
\end{center}
\caption{
Values of $M_B$ and $\Delta M_B$ and their percentage deviations for full pantheon data from the corresponding ones for the binned data. Here, the kernel covariance function is the squared exponential (SE) and the mean function is the $\Lambda$CDM.
}
\label{table:FuLL_Panth_vs_bin}
\end{table}


\bibliographystyle{apsrev4-1}
\bibliography{refs}

\end{document}